\RequirePackage{fix-cm}
\documentclass[natbib,smallextended]{svjour3} 

\smartqed  

\usepackage{graphicx}
\usepackage{aps-bibstyle}  

\usepackage{latexsym}
\usepackage{verbatim}
\usepackage{booktabs,caption}
\usepackage[flushleft]{threeparttable}
\usepackage[utf8]{inputenc}
\usepackage{longtable}
\usepackage{array}
\newcolumntype{H}{>{\setbox0=\hbox\bgroup}c<{\egroup}@{}}
\newcolumntype{Z}{>{\setbox0=\hbox\bgroup}c<{\egroup}@{\hspace*{-\tabcolsep}}}

\newcommand{\xraychap}{Saxton et al. 2021, X-ray Chapter}
\newcommand{\gammachap}{Zauderer et al. 2021, Gamma-ray Chapter}
\newcommand{\optchap}{van Velzen et al. 2021, Optical Chapter}

\newcommand{\hostchap}{French et al. 2021, Host Galaxies Chapter}
\newcommand{\ratechap}{Stone et al. 2020, Rates Chapter}

 \journalname{ISSI Book on TDEs}

\usepackage{ref_macros} 
\usepackage{graphicx}
\usepackage{amssymb}
\usepackage{xspace}
\usepackage[usenames,dvipsnames,svgnames,x11names,table]{xcolor}

\definecolor{Purple}{rgb}{0.5,0,0.5}
\definecolor{Orange}{rgb}{1,0.5,0}

\newcommand{\todo}{\ifmmode \color{red}\Huge{\bullet} \else {\color{red}{\Huge$\bullet$}}\fi}
\newcommand{  \heii     }{\ifmmode {\rm He}\,\textsc{ii} \else He\,\textsc{ii}\fi}
\newcommand{\HeIIop}{\ifmmode {\rm He}\,\textsc{ii}\,\lambda4686 \else He\,\textsc{ii}\,$\lambda4686$\fi}
\newcommand{  \OIII   }{\ifmmode $\left[{\rm O}\,\textsc{iii}\right]$\,$\lambda5007$ \else [O\,\textsc{iii}]\,$\lambda5007$\fi}
\newcommand{  \OIIIbf   }{\ifmmode {\rm O}\,\textsc{iii}\,$\lambda3133$ \else O\,\textsc{iii}\,$\lambda3133$\fi}
\newcommand{  \NIIIbf     }{\ifmmode {\rm N}\,\textsc{iii}\,$\lambda4640$ \else N\,\textsc{iii}\,$\lambda4640$\fi}
\newcommand{\MgII}{\ifmmode {\rm Mg}\,\textsc{ii}\,\lambda2800 \else Mg\,\textsc{ii}\,$\lambda2800$\fi}
\newcommand{  \niii     }{\ifmmode {\rm N}\,\textsc{iii} \else N\,\textsc{iii}\fi}

\graphicspath{{./}{figures/}}

\begin{document}

\title{Distinguishing Tidal Disruption Events from Impostors}

\titlerunning{Unusual and Impostor TDEs}        

\author{Ann Zabludoff \and Iair Arcavi \and Stephanie La Massa \and Hagai B. Perets \and Benny Trakhtenbrot \and B. Ashley Zauderer \and Katie Auchettl \and Jane L. Dai \and K. Decker French \and Tiara Hung \and  Erin Kara \and Giuseppe Lodato \and W. Peter Maksym \and Yujing Qin \and Enrico Ramirez-Ruiz \and Nathaniel Roth \and Jessie C. Runnoe \and Thomas Wevers
}

\authorrunning{Zabludoff et al.} 

\institute{A. Zabludoff (University of Arizona, aiz@arizona.edu)
\and I. Arcavi (Tel Aviv University) \and
S. La Massa (STScI) \and
H. Perets (Technion) \and
B. Trakhtenbrot (Tel Aviv University) \and
B. A. Zauderer (NSF) \and
K. Auchettl (University of Melbourne) \and
J. L. Dai (HKU) \and 
K.D. French (Carnegie Obs) \and
T. Hung (UCSC) \and
E. Kara (MIT) \and 
G. Lodato (Milan) \and 
W.P. Maksym (Harvard-Smithsonian CfA) \and 
Y. Qin (Arizona) \and
E. Ramirez-Ruiz (UCSC) \and
N. Roth (Maryland) \and
J. Runnoe (Vanderbilt) \and
T. Wevers (ESO)
}

\date{Received: date / Accepted: date}

\maketitle

\begin{abstract}


Recent claimed detections of tidal disruption events (TDEs) in multi-wavelength data have opened potential new windows into the evolution and properties of otherwise dormant supermassive black holes (SMBHs) in the centres of galaxies. At present, there are several dozen TDE candidates, which share some properties and differ in others.  The range in properties is broad enough to overlap other transient types, such as active galactic nuclei (AGN) and supernovae (SNe), which can make TDE classification ambiguous. A further complication is that ``TDE signatures"
have not been uniformly observed to similar sensitivities or even targeted across all candidates. This chapter reviews those events that are unusual relative to other TDEs, including the possibility of 
TDEs in pre-existing AGN, and
summarises those characteristics thought to best distinguish TDEs from continuously accreting AGN, strongly flaring AGN, SNe, and Gamma-Ray Bursts (GRBs), as well as other potential impostors like stellar collisions, ``micro-TDEs," and circumbinary accretion flows. We
conclude that multiple observables should be used to classify any one event as a TDE. We also consider the TDE candidate population as a whole, which, for certain host galaxy or SMBH characteristics, is distinguishable statistically from non-TDEs,
suggesting that at least some TDE candidates do in fact arise from SMBH-disrupted stars.
\end{abstract}

\newpage
\section{Introduction} \label{sec:Intro}

A TDE is a star disrupted by a SMBH.
The TDEs discussed in previous chapters span a range of observed characteristics, including candidates detected first or only in X-rays, optical, or UV light, and with or without broad H/He, coronal, or Bowen emission lines.
All are energetic transient events consistent with arising from galactic nuclei and are not known to have re-occurred. 

Yet there are non-TDE transients---large AGN flares, SNe near or projected on the nucleus---with potentially similar features. Even some ``normal," continuously accreting AGN may vary over timescales longer than some TDE candidates have so far been monitored. The range of TDE properties, which can overlap those of other transient types, suggests that there is no one observable that distinguishes TDEs unambiguously from impostors. Nor has the developing field of modelling TDE formation and emission found such a ``smoking-gun."  

For the time being, we must rely on the standards of jurisprudence and require a preponderance of evidence, a suite of distinguishing features. We also can employ statistical arguments to test the authenticity of at least some TDE detections. What are those features and arguments?

In this chapter, we first discuss those current TDE candidates whose properties are rare or new among claimed TDEs and examine the reasons to favour the TDE explanation. Then, for the remaining TDE candidates, we review those observables that are generally interpreted as TDE signatures, comparing them to what is known about continuously accreting AGN, AGN with strong flares arising from disk instabilities, SNe, and GRBs, as well as other potential impostors like stellar collisions, ``micro-TDEs," and circumbinary accretion flows. Lastly, we explore using TDE demographics, specifically
the projected offset relative to the galactic nucleus, the SMBH mass, 
and the stellar mass and star formation history of the host galaxy,
to distinguish them statistically from non-TDE sources.

%
%


\section{Unusual TDE Candidates}
\label{sec:oneoffs}


On-going time-domain surveys, and comprehensive follow-up campaigns, are continuously revealing new flares and transients in galactic nuclei.
Some objects initially thought to be TDEs
have been reclassified due to the subsequent detection 
of similar, non-TDE transients, e.g.,
F01004-2237 \citep{Tadhunter:2017a}, after the discoveries of OGLE17aaj \citep{Gromadzki2019_OGLE17aaj} and AT~2017bgt \cite[][see Section \ref{ref:flares_in_known_AGN}]{Trakhtenbrot2019_AT2017bgt}, illustrating the challenge of disentangling TDE emission from other peculiar nuclear transients.
Here we discuss unusual classes of objects that are still considered TDE candidates, but that are represented by only one or a handful of members.
The coming era of RubinObs/LSST and \emph{eROSITA}, when thousands of new TDE candidates will be identified, should reveal the true nature of these classes.

\subsection{PS1-11af}

PS1-11af is interpreted as a partial TDE (i.e., a disruption of the envelope of the star, leaving the core intact) by \citet{Chornock:2014a}. Absorption features became apparent in the UV spectrum 24 days post flare that bear a similarity to P-Cygni troughs observed in SNe. Yet the apparent velocities of these features ($\sim$13,000 km s$^{-1}$) are too high for material in homologous expansion near a SN photosphere. Furthermore, fits to the SED with a blackbody model show that the radius of the emitting ejecta does not increase and the temperature does not decrease, as would be expected for SN evolution. PS1-11af has a blue colour that evolves weakly over time, a common characteristic of optically-detected TDEs \citep[e.g.,][]{van-Velzen:2011a, Gezari:2012a}.
The amount of accreted mass needed to power the observed luminosity is low ($\sim$ 0.002$M_{\odot}$), which \citet{Chornock:2014a} ascribe to the partial stellar tidal disruption. In this scenario, optical photons would be reprocessed from the accretion disk to higher (X-ray) energies \citep[e.g.,][]{Strubbe:2009a, Guillochon:2013a}, which requires contemporaneous X-ray coverage to confirm this hypothesis. Lacking the observations at this energy range, \citet{Chornock:2014a} can only propose that a partial stellar tidal disruption explains the available data, but the picture remains necessarily incomplete, underscoring the importance of simultaneous multi-wavelength coverage in unveiling the physics driving transient phenomena.

\subsection{ASASSN-15lh}
ASASSN-15lh was first interpreted as the most luminous SN ever detected \citep{Dong:2016}. 
Its optical spectrum is devoid of hydrogen and helium features and contains broad absorption features between 3000 - 4100 \AA. One such feature is attributed to OII~$\lambda$4100, which is also seen in hydrogen-poor superluminous SNe \citep[e.g.,][]{quimby2011}. However, an additional OII~$\lambda$4400 feature observed in SNe
is not present.
This discrepancy, together with the location of ASASSN-15lh in the center of a non-starforming massive galaxy, prompted \citet{Leloudas:2016a} to consider ASASSN-15lh as a TDE (see also \citealt{Kreuhler2018, van-Velzen:2018a, 2020MNRAS.497L..13M}). 
The inferred SMBH mass from simple galactic scalings exceeds 10$^{8}$ $M_{\odot}$, implying that a solar-mass, solar-radius star would be swallowed whole rather than disrupted, but \citet{Leloudas:2016a} point out that a spinning SMBH would tidally disrupt the star and produce a transient (see also \citealt{Margutti:2017a}).
As a result, ASASSN-15lh, if indeed a TDE, could then be used to infer the SMBH spin, a property that is challenging to constrain with most other methods. 

Another peculiar property of ASASSN-15lh is its double-peaked UV light curve (its optical light curve is single-peaked). This bimodality, also observed
in the TDE candidate AT2018fyk (\citealt{Wevers2019}, see below),
is unusual and difficult to explain under both the SN and TDE interpretations. \citet{Leloudas:2016a} propose that the first peak is powered by circularisation of the debris, while the second is from accretion; the timescales are roughly consistent with those expected from a spinning SMBH.
Alternatively, \citet{2018MNRAS.474.3857C} and \citet{2018MNRAS.476.5312V} argue that both the double peaked light curve and the apparent high SMBH mass can be explained if the TDE is due to the secondary in a SMBH binary system. Still, the nature of this TDE candidate is disputable;
\citet{2017MNRAS.466.1428G} argue that the evolution of ASASSN-15lh's photospheric radius, its radiated energy, and the implied event rate are all more consistent with those of H-poor superluminous SNe than TDEs.

\subsection{PS16dtm}
\label{sec:ps16dtm}

The transient PS16dtm, discovered in a Narrow Line Seyfert 1 (NLSy1)
galaxy,
was interpreted as a TDE rather than as a SN or arising from intrinsic AGN variability \citep{Blanchard:2017a}. The light curve exhibits no colour evolution during the $\sim$100 day plateau,
similar to other optical TDEs (see \optchap). The optical spectrum has traits similar to NLSy1s, with Balmer and multi-component FeII emission lines, further arguing against a SN interpretation. \citet{Blanchard:2017a} rule out AGN variability, given the two orders-of-magnitude increase in optical/UV flux within $\sim$50 days and the decrease in X-ray flux after the optical/UV flare; they ascribe this behaviour to obscuration of the pre-existing AGN X-ray corona by the stellar debris disk formed by the disrupted star. The 
rise in the light curve is followed by a plateau at roughly the Eddingtion luminosity inferred for the SMBH and then a decline. 

While this source is a strong TDE candidate, its spectrum, which is AGN-like, is very different than those of many optical TDEs (see \optchap).
Indeed, \citet{Moriya2017} point out that PS16dtm's flare can be explained by AGN activity: increases in the accretion disk luminosity can spur radiatively driven winds that cause shock waves to propagate within the BLR. Interactions between the shocks and BLR clouds can convert kinetic energy of the ejecta into radiation, producing transient luminosities and timescales that match those observed in PS16dtm.

\subsection{AT2018fyk}
AT2018fyk is a TDE candidate with a photometric UV/optical evolution remarkably similar to ASASSN--15lh, i.e., with a secondary maximum in its light curve. \citet{Margutti:2017a}
ascribe the second peak in ASASSN--15lh's
light curve to temporal evolution in the opacity of the ejecta, which allows UV radiation to escape and produce the secondary maximum. They point out that the observations are consistent with a spinning black hole disrupting a main-sequence star as a trigger for the ASASSN-15lh flare. Despite the similarities in light curve evolution, 
the timescales for AT2018fyk are significantly shorter than for ASASSN--15lh, and high amplitude, erratic X-ray variability is observed for AT2018fyk from the early phases. 

AT2018fyk's optical-to-X-ray luminosity ratio (L$_{\rm opt}$/L$_{\rm X}$) evolves like that of ASASSN--15oi, suggesting that similar physical processes are at play.
\citet{Wevers2019} argue that both the L$_{\rm opt}$/L$_{\rm X}$ evolution and the secondary maximum in the UV/optical light curve can be explained as a tidal disruption with a relativistic pericenter, as \citet{Leloudas:2016a} suggested led to the
double-humped light curve in 
ASASSN--15lh. A relativistic pericenter favours disk formation on short ($\sim$ months) timescales, compared with the typical timescale of $\sim$years \citep{vVelzen2019}. \citet{Gezari:2017a} note that the formation of an accretion disk on similarly short timescales may explain the peculiar L$_{\rm opt}$/L$_{\rm X}$ observed in ASASSN--15oi. 

Another peculiarity of AT2018fyk
is the apparent decoupling of the X-ray from the UV/optical emission, about 80 days after peak. This is reminiscent of the late time X-ray detection in ASASSN--15lh, while the UV/optical lightcurve steadily declines. \citet{Margutti:2017a} postulate that in ASASSN--15lh, the X-ray emission may not in fact be related to the transient and may arise instead from the host galaxy nucleus, which would favour an interpretation that the flare was caused by a stellar explosion rather than a TDE. Should the X-ray emission be due to a TDE from a massive spinning black hole, the X-ray emission would fade over time. A similar observational test can be brought to bear on AT2018fyk by monitoring its X-ray emission over the time span of years.


In the optical spectra of AT2018fyk, \citet{Wevers2019} detect low ionisation potential Fe\,\textsc{ii} emission lines  like those identified in ASASSN--15oi at late times. These lines are thought to form in dense, optically thick gas in an accretion disk-like structure, favouring the rapid disk formation scenario. These lines are observed frequently in high accretion rate NLSy1s, suggesting that the physical conditions in some TDEs and AGN are similar. Arguments against the AGN interpretation for this event include the absence of galactic (narrow or broad) emission lines, a pre-flare X-ray non-detection, and IR colours consistent with a quiescent galaxy.

\subsection{Summary}

There is not one selection mechanism that can be used to distinguish among potential explanations for transient events. Classification instead relies on the preponderance of evidence and may still not be definitive (e.g., ASASSN-15lh). Some transient phenomena have characteristics of both SNe and TDEs (e.g., PS1-11af), AGN and TDEs (e.g., AT2018fyk), or are hosted in known active galaxies (e.g., PS16dtm), requiring care in distinguishing among flares in a pre-existing accretion disk within the high variability tail of the AGN population, the tidal disruption of a star in the vicinity of an already active black hole, or a supernova in the centre of a galaxy.

The next several sections discuss how we might differentiate TDEs from the signatures of AGN (and strongly flaring AGN), SNe, GRBs, and other potential impostors, based on photometric and spectroscopic clues as well as statistical arguments.

\section{Distinguishing TDEs from AGN} 
\label{sec:TDEvAGN}

Dozens of TDEs are now claimed to have been detected. The uncertainty in this number reflects the inhomogeneity of TDE definitions, incompleteness in the TDE observables, and the lack of a unifying theoretical framework.  
TDEs should differ from AGN in the details of their accretion, i.e., the disruption of a single star leading to the quick, inside-out formation of
a small, initially inclined
disk (or flow) that then disappears on a shorter timescale than typical of 
the more continuous nature of AGN accretion. 

Yet there is much we do not know about AGN variability, particularly
about the extremes of continuous variability and
about instabilities in the accretion disk that may produce transient flaring. The discovery of new variable AGN classes such as hyper-variable and ``changing-look" AGN, which show dramatic weakening and/or strengthening in their broad Balmer emission lines, complicates efforts to identify TDEs unambiguously. Even long-term AGN variability may be a problem; the relevant timescales for TDEs with evolved stellar progenitors (which are not the main focus here) could be far longer
than for main sequence stars.
Much theoretical work remains to predict TDE observational signatures and to ascertain which, if any, are unique to TDEs. 

In the following discussions, we consider what may distinguish TDEs from continuously accreting AGN, including those that are highly variable, and, more problematically, from the flaring caused by AGN disk instabilities.
An even bigger challenge is presented by a new class of events---combining TDE- \emph{and} AGN-like observables---that may arise when a TDE occurs in a pre-existing AGN \cite[e.g.,][]{Merloni:2015a,Chan19,Ricci2020_1ES}.
%
In \S\ref{ref:flares_in_known_AGN},
we briefly discuss the few such objects detected to date,
reflecting our limited knowledge at the time of this writing.

The guidelines presented below tend to err on the conservative side: we are more interested here in purity than completeness in TDE classification.  
As a result, TDEs that occur in galaxies with even mild signs of nuclear activity, e.g., Seyfert-like emission line ratios, otherwise strong \OIII\ emission, or persistent archival X-ray emission, would be excluded by our criteria. 
Yet counting hybrid systems will be critical in building complete and unbiased TDE samples.

\subsection{TDEs versus Continuous AGN} \label{sec:Continuous_AGN}

The obvious difference between TDEs and steady-state AGN is that TDEs are fundamentally transient phenomena, transitioning from quiescence to near-Eddington luminosities in a few weeks, and then back to quiescence within a few years to even decades \citep{Rees:1988a,Stone:2013a,vVelzen2019}. This opens the possibility of monitoring several state transitions in the accretion flow, which goes from near- (or super-) Eddington to sub-Eddington and eventually becomes radiatively inefficient at low accretion rates \citep{Jonker2020}. Long term monitoring of TDEs should be pursued to reveal the details of such state transitions. 

Debris disks from the tidal disruption of main sequence stars are very compact, as the star is disrupted near the SMBH with low angular momentum. If the stellar debris circularises efficiently, the size of the compact disk formed is twice the tidal disruption radius, or $\sim 10-100 R_{\rm g}$ where $R_g = G M_{BH}/c^2$. In contrast, AGN disks are expected to be much more extended, as gas is supplied from farther distances \citep[e.g.,][and references therein]{Alexander2012}. 

Another difference between TDE and stable AGN is that TDE disks may be fed at super-Eddington rates, while (low-redshift) AGN are usually considered to be accreting at sub-Eddington levels. As a result, TDEs and these AGN would have different disk structures, i.e.,
the super-Eddington TDE disks would be geometrically and
optically thick and produce optically thick winds \citep{Strubbe:2009a,Lodato:2011a, Dai18}.
As the accretion rate drops with fallback rate, the disk and wind densities will also decrease, lowering the electron scattering opacity. This behaviour can explain the narrowing of the TDE hydrogen and \HeIIop\ emission lines with decreasing luminosity described below, as the line width in a scattering dominated medium scales with the opacity \citep{Roth18}.  
The higher \heii/H$\alpha$ ratio observed in TDEs (also discussed below) can arise from the higher accretion levels and inner disk temperatures compared to AGN, although detailed modelling on using this ratio to directly probe the disk structure is still lacking. 

The absence of hard X-ray emission in the TDEs observed so far, compared with the X-ray power-law spectrum with $\Gamma \sim 1.9$ common to AGN (as discussed below), suggests fundamental differences in the disk corona. Possible explanations include: 1) the typical duration of AGN accretion is much longer than a TDE lasts, so the corona forms only for AGN; 2) the magnetic field strength and configuration is different in TDEs than in AGN, leading to less efficient coronal production.

\subsubsection{Summary of Observable Distinctions}

\label{sec:summary_guidelines_AGN}

Some continuously accreting, but variable, AGN may be identified as new, blue, and/or X-ray detected nuclear transients and thus misclassified as TDE candidates. Consequently, it is essential that we consider the breadth of known AGN properties---light curves, colors, spectral shapes and lines, and variability, across optical, UV, and X-ray wavelengths---in defining criteria that may distinguish TDEs. In what follows, we discuss possible criteria and demonstrate how they may be used to assess the data for the two best-studied TDE candidates, ASASSN-14li and ASASSN-15oi.
Given that strongly flaring AGN may have properties distinct from the continuous AGN population, posing a different and perhaps greater challenge to TDE classification, we discuss other appropriate strategies in \ref{sec:Flaring_AGN}.

Features that may favour a TDE over other AGN activity include:

\begin{enumerate}

\item steeper (month-long) and brighter (change of several magnitudes) rise in optical/UV flux;\label{prop_list:fast_rise}

\item relatively narrow luminosity peak, with characteristic timescale of months;\label{prop_list:narrow_peak}

\item smooth, power-law decline in light curve, sometimes following a $t^{-5/3}$ trend;\label{prop_list:pl_decline}

\item $\sim$0.2 mag bluer in $g$-$r$ around peak emission;\label{prop_list:bluer}

\item hot, constant $T \sim 2$-$4 \times 10^4$ K blackbody in optical/UV emission;\label{prop_list:UVopt_BB}

\item absent to weak \OIII\ emission, and narrow emission line ratios suggestive of star formation rather than AGN photoionisation;
\label{prop_list:nlr}

\item very broad ($>$15,000 km s$^{-1}$) \HeIIop\ and Balmer optical emission lines that narrow as they weaken;\label{prop_list:blr_evo}

\item luminous \HeIIop\ line emission, with \heii/H$\alpha$ flux ratio $\gtrsim1$;\label{prop_list:heii_strong}

\item weak, or even absent, \MgII\ line emission;\label{prop_list:mgii_weak}

\item softer X-ray spectrum, in terms of photon index ($\Gamma \geq 3$) and/or prominence of low-temperature emission component ($kT_{\rm bb}$ = 0.04--0.12 keV; see the \xraychap)\label{prop_list:xray_soft}

\item less rapid ($>$ hours) X-ray variability;\label{prop_list:xray_var} 

\item no recurrence of transient behaviour\footnote{One possible exception is if a TDE occurs in a binary SMBH. In this case, the TDE may be perceived as recurring transient behaviour when the X-ray light dims due to the interaction with the second SMBH.}.\label{prop_list:no_recurrence} 

\end{enumerate}
 
No TDE candidate observed to date has been shown to possess all these features. Indeed, depending on the conditions, some TDEs may not generate certain features.
Those TDEs with the most features above, e.g., ASASSN-14li and ASASSN-15oi, are considered the strongest TDE candidates. 
We discuss these two events in light of the list given above, before addressing each of the listed features in more detail.

\paragraph{ASASSN-14li:\\}
\smallskip

Here is how the optical, UV, and X-ray observations of ASASSN-14li map to the criteria listed above. 
1) It displayed a $\Delta$UVW2 (\emph{Swift}) of -4.1 and a $\Delta g$ of only -0.4, where these variations in magnitudes were measured with respect to the host galaxy pre-flare archival measurements  \citep{Holoien:2016b};  
2) The peak of this event was not observed; 
3) The UV/optical decline over the first six months of monitoring was initially fit with an exponential \citep{Holoien:2016b}, but this emission was later fit with a $t^{-5/3}$ decline over a longer time interval of approximately 250 days. After this, the UV/optical light curve levelled off to a more shallow decline \citep{2017MNRAS.466.4904B}; 
4) The $g-r$ colour was roughly 0.4 mag during early monitoring, but was highly affected by host contamination. The event was still quite blue, with UVM2 brighter than U (\emph{Swift}) by a difference exceeding 0.5 mag, for at least the first 100 days of monitoring \citep{Holoien:2016a}; 
5) The optical/UV continuum can be fit with a blackbody with $T \approx 3.5 \times 10^4$ K, and this temperature remained almost unchanged for the first 175 days of monitoring \citep{Hung2017}.

Furthermore, 6) optical spectra indicate \OIII/H$\beta$ $\ll 1$ \citep[][and see also Figure~\ref{fig:H_He_lines} in this chapter]{Holoien:2016b}; 
7) The optical emission lines initially showed broad wings with widths of $\sim 10,000$ km s$^{-1}$, although these widths narrowed significantly, with only a narrow component of width $\sim 1500$ km s$^{-1}$ after 100 days \citep{Holoien:2016b}.  The optical emission lines in the later spectra from this event are among the narrowest seen in TDEs; 
8) While \heii/H$\alpha$ varies, this ratio was $> 1$ for several epochs and generally exceeded 0.5 \citep{Hung2017};
9) No \MgII\ emission was seen in UV spectroscopy \citep{Cenko:2016a}.

Lastly, 10) the X-ray spectrum was soft and could be fit with a $kT = 51$ eV blackbody \citep{Miller:2015b}; 
11) X-ray variability \emph{was} detected in this event: a stable quasi-periodic oscillation of roughly 131 seconds was identified \citep{Pasham2019}; 
12) No recurrence has been observed. 

In summary, 14li meets most of the criteria for distinguishing a TDE from other AGN activity. The most prominent exceptions are for criteria 4 and 11: the earliest (closest to peak) $g - r$ measurement was only 0.2 (but this might be due to host contamination), and the event exhibited rapid X-ray variability in the form of a QPO. Additionally, since the peak was not observed, it is not possible to assess whether the characteristic rise and fall timescales are on the order of months  (criterion 2). Finally, optical emission lines were somewhat narrow compared to other putative TDEs (criterion 7), although these lines did narrow over time in a manner that seems characteristic of TDEs.


\paragraph{ASASSN-15oi:\\}
\smallskip

Next, we consider ASASSN-15oi. 
1) It displayed a $\Delta$UVW2 ({\it Swift}) of -6.8 and a $\Delta V$ of only -1.2  \citep{Holoien:2016a}, where once again these variations in magnitudes were measured with respect to the host  galaxy  pre-flare  archival  measurements; 
2) The peak of this event was not observed; 
3) A UV/optical decline similar to $t^{-5/3}$ could be fit to the first 100 days of observation, before the flux at these wavelengths dropped precipitously. While the flux initially declined steadily in all bands, the inferred bolometric flux (from a thermal fit to the optical/UV data) remained steady for approximately the first 50 days before entering a decline \citep{2018MNRAS.480.5689H}. As with ASASSN-14li, the initial UV/optical decline could alternatively be fit with an exponential \citep{Holoien:2016a};
4) The $g$-$r$ measurement was not published, but UVM2 was brighter than U ({\it Swift}) by at least 1.0 mag during the first 50 days of monitoring (10 - 60 days post-discovery) and by at least 0.5 mag for 40 days after that  \citep{Holoien:2016a}. 
5) During roughly the first 15 days of monitoring, the optical/UV continuum could be fit with a blackbody of roughly $T\sim 2 \times 10^4$ K. The inferred temperature \emph{increased} to about $4 \times 10^4 K$ over the next 15 days and stayed at that temperature for at least the next 70 days \citep{Hung2017}. 

In addition, 6) the [O\,{\sc iii}] lines do not appear prominently in the spectra, although there is an unidentified broad feature near 5000 \AA, which can be seen at 21 days post-discovery in the host-subtracted spectrum. 
Likewise, H$\beta$ does not appear prominently, although it may be blended with \heii\ in the earliest spectrum \citep[][and see also Figure~\ref{fig:H_He_lines} in this chapter]{Holoien:2016a}.  
7) The \HeIIop\ emission line had a width (FWHM) of roughly 20,000 km s$^{-1}$ in a spectrum taken seven days after discovery, which narrowed to approximately 10,000 km s$^{-1}$ at 21 days \citep{Holoien:2016a}. 
8) While \heii\ appears prominently in the spectrum, no clear detection of any hydrogen lines can be made \citep{Holoien:2016a}; 
9) No Mg\,\textsc{ii} emission was seen in UV spectra (Alexander Dittmann et al., in prep.). 

Finally, 10) the X-ray spectra were soft and could be fit with blackbodies with $kT \sim 40 - 50$ eV; 
11) No rapid X-ray variability has been reported; 
12) No recurrence has been observed. 

In summary, 15oi meets nearly all the aforementioned criteria to distinguish a TDE from other AGN activity, except for those specific to the peak of the light curve (criteria 2 and 4), which was not captured for this event.


\subsubsection{UV to Optical Light Curve}

The light curves of TDE candidates are characterised by a dramatic increase in optical and UV luminosity, with an observed variability of over three magnitudes  \citep[e.g.,][]{van-Velzen:2011a},
a narrow luminosity peak spanning a timescale of months \citep[e.g.,][]{Gezari:2009a,Guillochon:2013a},
and a smooth power-law decline, which sometimes follows $t^{-5/3}$, the predicted mass fall-back rate \citep{Rees:1988a,Phinney1989_TDEs}.

\begin{figure*}
\centering
\includegraphics[width=0.48\textwidth]{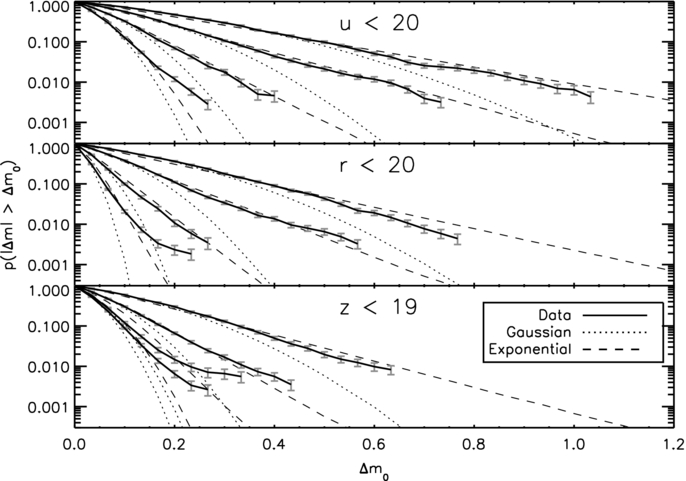}
\hfill
\includegraphics[width=0.47\textwidth]{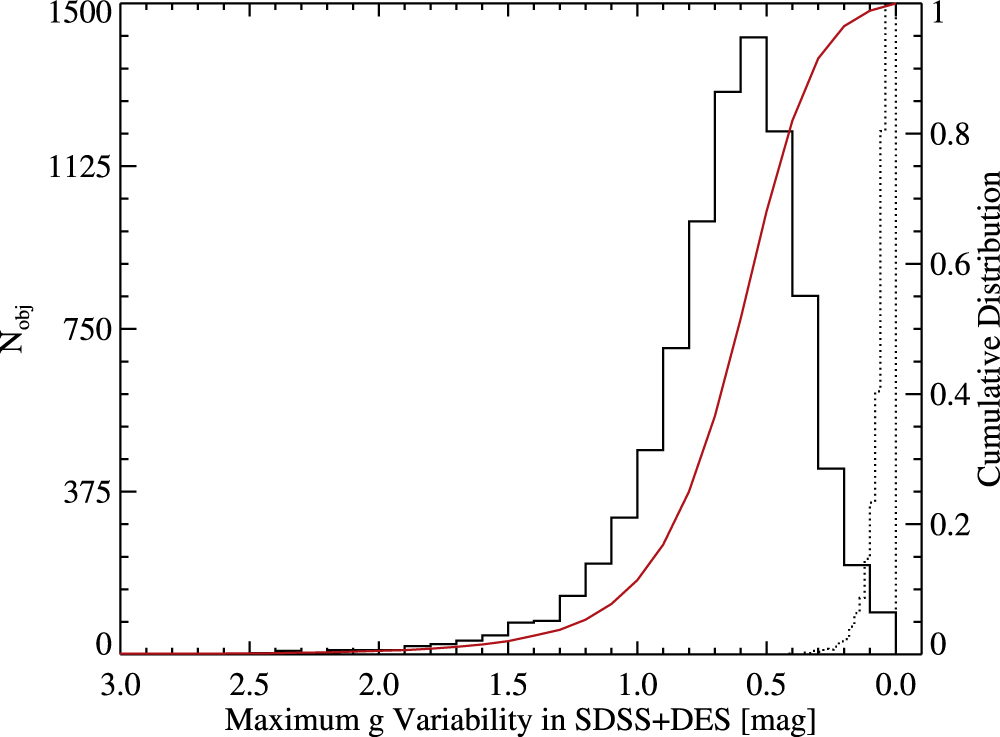}\\
\vspace{0.5cm}
\includegraphics[width=0.85\textwidth]{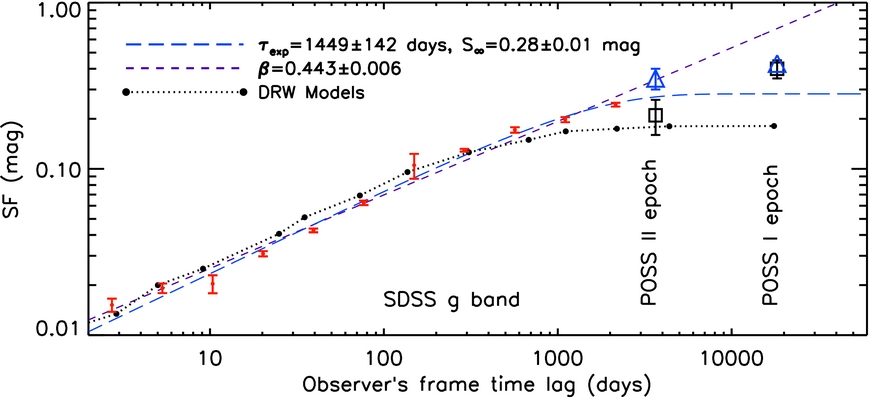}
\caption{Optical variability of normal, persistent (non-flaring) AGN.
{\it Top-Left:} Distribution of UV/optical flux variations of quasars in the SDSS+POSS study of \citet{MacLeod2012_SDSS_POSS_DRW}. 
In each panel, different solid lines trace quasar variability on timescales spanning 1-30 days, 50-150 days, 200-400 days, and 1400-3000 days---the former being relevant to the rise-time of most TDEs.
{\it Top-Right:} Distribution of optical flux variations in DES+SDSS broad-line AGN, over a period of $\sim$15 years (adopted from \citealt{Rumbaugh2018_DES_EVQs}). 
The cumulative distribution function (red line) indicates that only $\sim$10\% of AGN show $|\Delta mag| > 1$ over this long period and are claimed to be highly variable on all timescales.
{\it Bottom:} AGN variability on multiple timescales described through a structure function (SF), a measure of the {\it rms} variability of an AGN sample over any given time separation. 
This example (again from  \citealt{MacLeod2012_SDSS_POSS_DRW}) shows SDSS measurements over several years (red points) and combines them with POSS data for the longer-timescale measurements (large symbols). 
The different lines are phenomenological fits to the data.
On timescales of months,
AGN typically vary by $<0.1$ mag, while TDEs show up as $\gtrsim1$ mag transients (i.e., corresponding to ${\rm SF}\gtrsim1$ mag over $<$ 100 days).
Thus, normal, persistent (non-flaring) AGN essentially never show the month-long sharp optical flux increase seen in TDEs.
}
\label{fig:AGN_var_stats}
\end{figure*}

In comparison, the light curves of variable AGN are dominated by more stochastic variability that lacks such dramatic increases in brightness.
On timescales of months, the variability of the continuous AGN population rarely exceeds $0.1$ mag \cite[Fig.~\ref{fig:AGN_var_stats}; see, e.g.,][and references therein]{VandenBerk2004,macleod2010,van-Velzen:2011a,MacLeod2012_SDSS_POSS_DRW,Caplar2017_PTF_QSOs}.
Although this {\it typical} AGN optical variability amplitude increases towards longer timescales, it does not exceed $\Delta{\rm mag}\sim1$, even over decades.
Indeed, only the most extreme tail of AGN variability distribution, with few sources in wide-field surveys, reaches $\Delta {\rm mag} \sim 2$ \citep[top panels of Fig.~\ref{fig:AGN_var_stats}; see, e.g.,][]{MacLeod2012_SDSS_POSS_DRW,Graham17,Rumbaugh2018_DES_EVQs}.
Likewise, AGN typically do not show smooth and steady variability structure, such as the power-law decline 
seen in TDEs.

Some observed changing-look AGN stay at their peak optical luminosity for years \citep[e.g.,][]{runnoe}. 
While there are TDE candidates detected in X-rays over a similar timescale \citep{Lin:2017a, Jonker2020}, the optical flare fades much more quickly.
Furthermore, although the decays in some changing-look AGN light curves approximate a $t^{-5/3}$ decline \cite[e.g.,][]{Merloni:2015a,Trakhtenbrot2019_1ES1927} or permit a $t^{-5/3}$ solution \citep[e.g.,][]{runnoe}, perhaps suggesting triggering by TDEs, many of these extremely variable AGN wane
differently \citep{ruan} and/or lack the smooth decline 
expected from the fallback of debris from a TDE \citep{Gezari:2017a}.

The evolution of optical colours can also be used as a selection criterion for TDEs in ground-based optical surveys. Optically-discovered TDEs are characterised by a long-lasting blue continuum that resembles a blackbody of a few $\times 10^4$ K. Unlike SNe, which typically undergo significant colour evolution over a few weeks, TDEs and AGN can keep a constant colour for a longer period of time ($\sim$years). The $g$-$r$ colour of TDEs at peak emission ($<-$0.2 mag) is typically bluer than for AGN \citep[$>-0.1$ mag; Fig~12 in][]{van-Velzen:2011a}. 
The observed bluer-when-brighter trend for AGN is consistent with the simple scenario of a geometrically-thin, optically-thick (i.e., Shakura-Sunyaev like) disk with variable accretion rates \citep[e.g.,][]{pereyra2006,hung2016}. If the observed AGN flare is an intrinsic property of the accretion disk, the classic thin disk model would predict a $g$-$r$ of $\sim-$0.1 mag. Although intrinsic extinction in AGN is hard to estimate, it will only make the $g$-$r$ colour in AGN flares redder than the predicted value, separating them further from TDEs in the optical colour space.

\subsubsection{Optical Spectrum}

A few key features in the optical spectra of TDE candidates can be used to differentiate them from persistent AGN. 
\cite{VandenBerk2001} and the references therein provide more 
information about the typical UV/optical spectral energy distribution (SED) and emission lines in AGN. 
For complementary composite spectra in the NIR and UV, see
\cite{Glikman2006_NIR_SED} and \cite{Shull2012}, respectively.

\paragraph{Hot, Constant Blackbody Continuum:\\}
\smallskip

The continuum colour variability also differs between AGN and TDEs.  Quasars are known to be bluer when brighter \citep[e.g.,][]{macleod2010,ruan2014}.  Although there is substantial scatter in this relationship, it provides additional leverage in identifying TDEs where no colour evolution is observed due to the constant black body temperature that produces the optical/UV continuum emission.

\paragraph{Weak [O\,\textsc{iii}] Line:\\}
\smallskip

The AGN narrow line region (NLR) is primarily ionised by the accreting black hole, and can span scales of order $\sim0.1$-1 kpc, with some dependence on the AGN continuum luminosity \cite[e.g.,][]{Bennert2002_NLR_RL,Bennert,Mor2009, Hainline2013,Hainline2014}. The \OIII\ emission line is one of the most prominent in the NLR and has been used both to map out the size of the NLR \citep[e.g.,][]{Schmitt2003a,Schmitt2003b} and as a proxy of the intrinsic (bolometric) AGN luminosity \citep{kauffmann,heckman,lamassa2010,pennell2017}. 
Due to the much larger size scale of the \OIII\ emitting region (and thus of the NLR) compared with the broad line region (BLR), this line responds slower to the change in the ionising continuum than the broad emission lines \cite[i.e., $\gg100$ years; see, e.g.,][]{Peterson2013_N5548_NLR_RM}.

TDEs tend to have weak to no \OIII\ emission, with [O\,{\sc iii}]/H$\beta$ and [N\,{\sc ii}]/H$\alpha$ emission line ratios consistent with photoionisation from star formation or LINER-like activity on the BPT \citep{bpt, kewley} diagram. Even TDE hosts with Seyfert-like line ratios, e.g., ASASSN-14ae \citep{2017ApJ...835..176F}, ASASSN-14li \citep{2017ApJ...835..176F}, and iPTF16fnl \citep{Onori2019}, generally have weak line strengths that would classify them as LINER-like on a WHAN \citep{2010IAUS..267...65C} diagram. We explore the range of AGN signatures in TDE host galaxies in the \hostchap.

On the other hand, galaxies with significant nuclear photometric and spectroscopic variability and strong [O\,{\sc iii}] emission are more likely to be hosting highly variable AGN, especially if the emission line ratios are within the Seyfert region of the BPT diagram. The emission line ratios of many changing-look AGN are generally consistent with those of AGN \citep[e.g.,][]{runnoe,ruan}. Thus, our guideline 6 in \S\ref{sec:summary_guidelines_AGN} would exclude most strong AGN and most changing-look AGN. We note that, while some luminous quasars have relatively weak [O\,{\sc iii}] emission \cite[e.g.,][]{Netzer2004_NLR}, their UV/optical continuum luminosities tend to be far higher than those of TDEs and could be used as a discriminant.

What do we miss with this conservative cut, where we have prioritised TDE sample purity over completeness by selecting against strong [O\,{\sc iii}] emission? We would neglect, for example, the TDE candidate PS16dtm \citep{Blanchard:2017a}, whose host is a NLSy1. Also excluded would be transients in the 2017-bgt class (\citealt{Trakhtenbrot2019_AT2017bgt};
see below), whose nature is unclear. The presence of some activity in a galaxy does not eliminate \emph{a priori} the possibility of a TDE, although caution should be used in such cases.

\begin{figure*}	
\begin{center}
\includegraphics[width=0.8\textwidth]{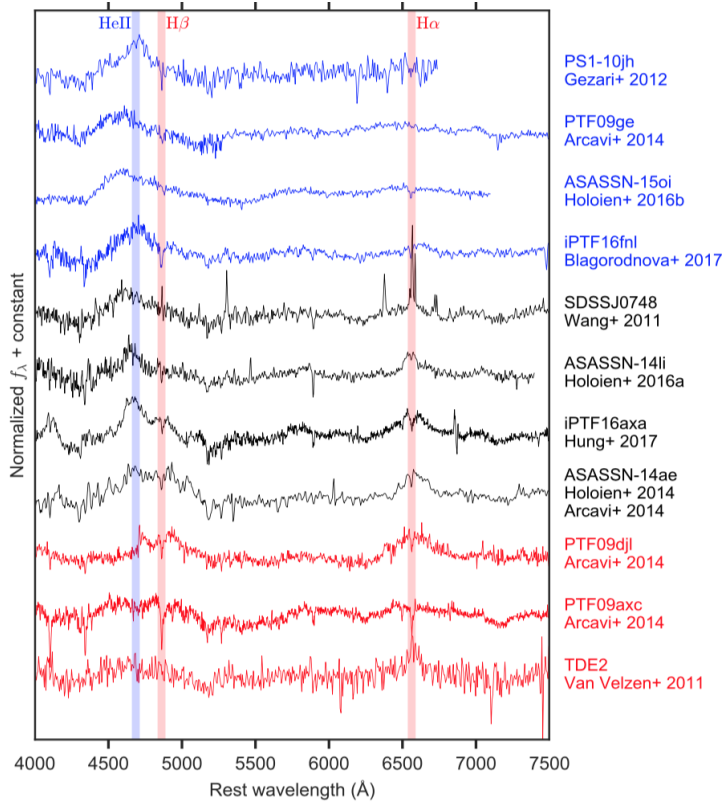}
\caption{Continuum-subtracted optical spectra of UV/optical-bright TDEs, most of which show He~\textsc{ii} line emission of comparable luminosity and FWHM to H$\alpha$.}
\label{fig:H_He_lines}
\end{center}
\end{figure*}

\paragraph{Broad, Narrowing Balmer and He~\textsc{ii} Lines:\\}
\smallskip

\begin{figure}	
\includegraphics[width=0.49\textwidth]{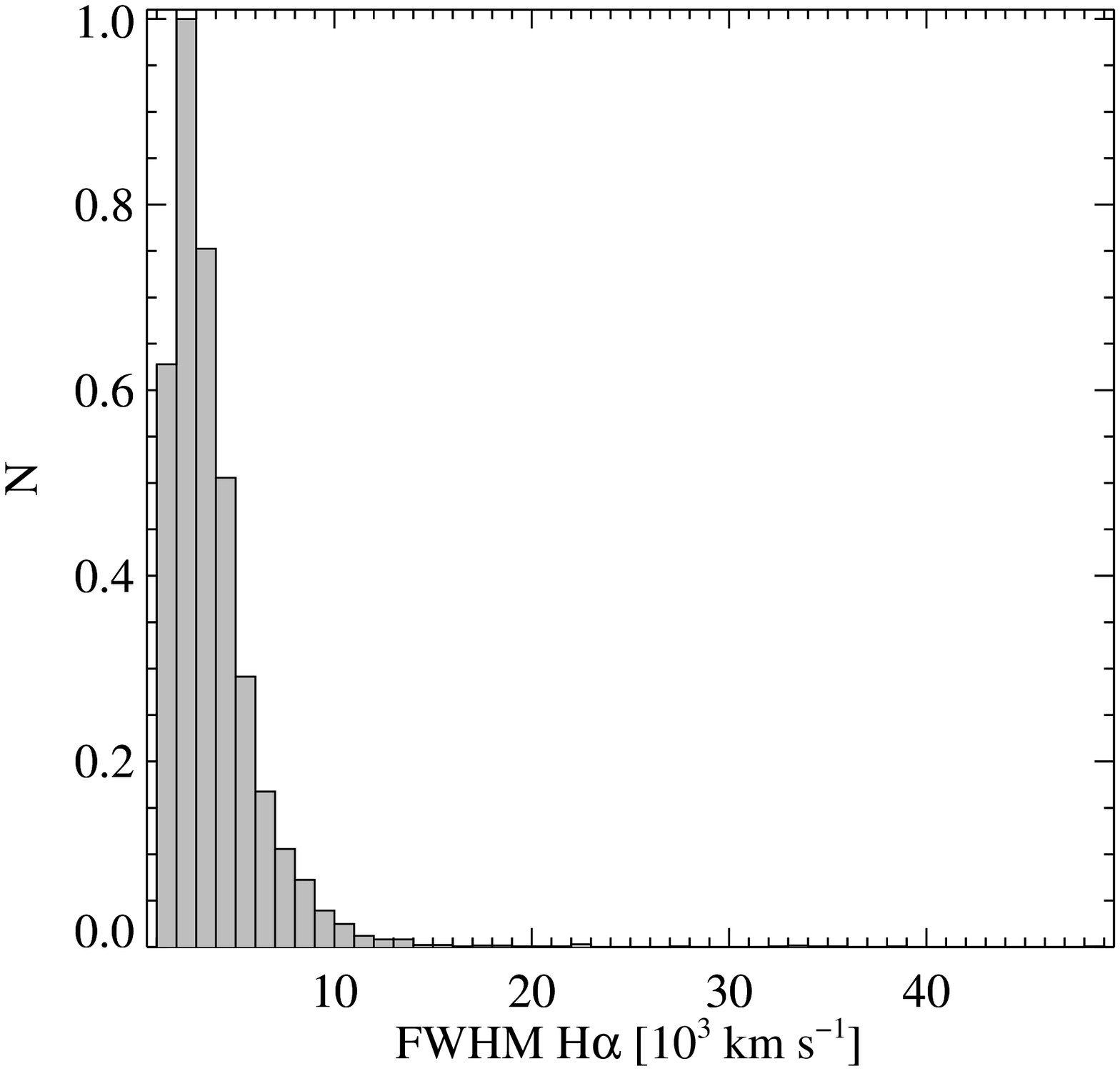}
\includegraphics[width=0.49\textwidth]{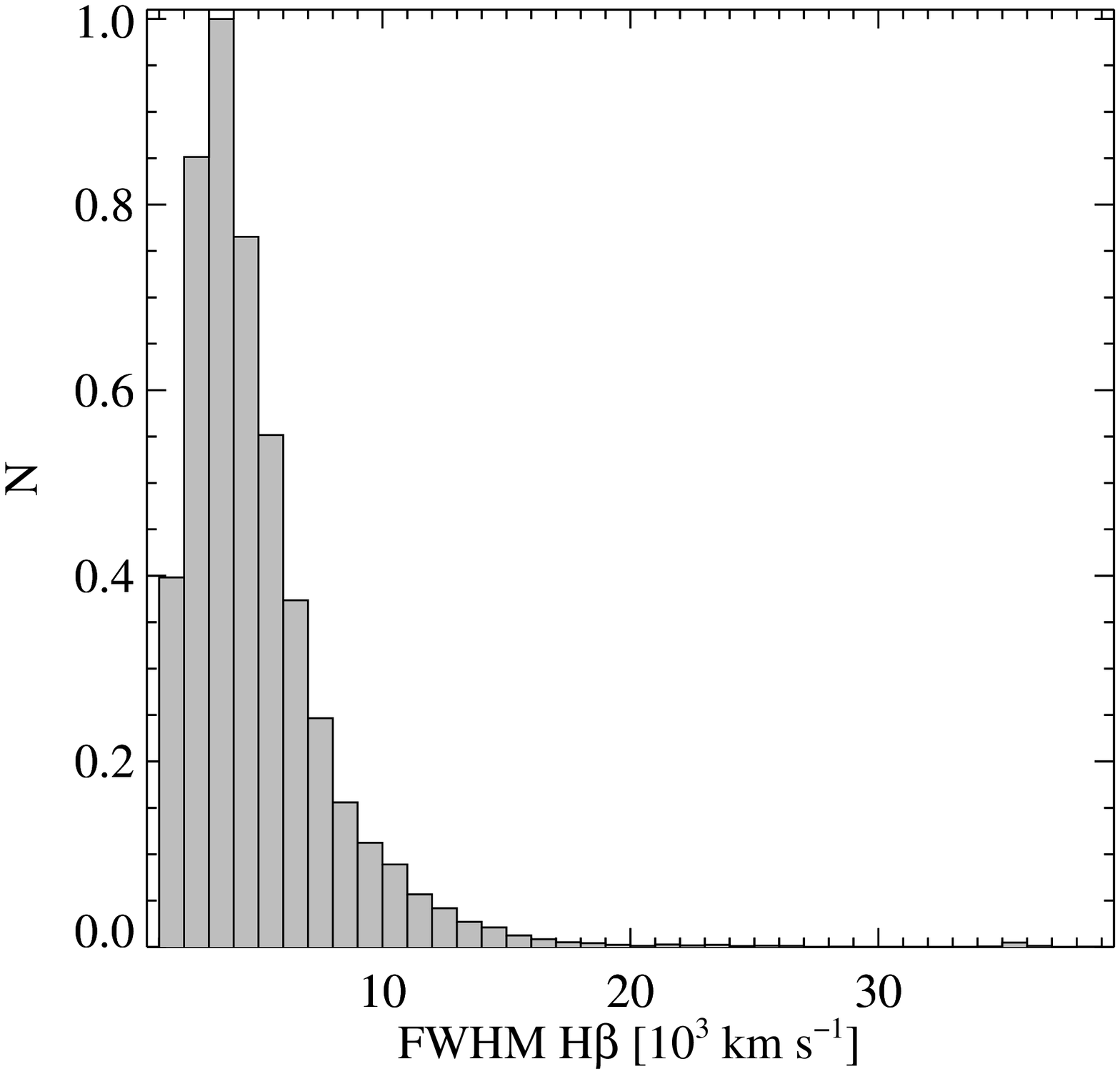}
\caption{The FWHM distributions for the H$\alpha$ and H$\beta$ broad emission lines in SDSS DR7 quasars, taken from the spectral decompositions of \cite{Shen_dr7_cat_2011}.
Only a small fraction of sources have Balmer lines that are broader than $\sim 15,000$~km~s$^{-1}$.
Given that such extremely broad Balmer lines are often seen in TDEs, Balmer line width can serve as a discriminant between TDE candidates and persistent AGN.
}
\label{fig:agnfwhm}
\end{figure}

The emission line velocity widths and their time-dependent changes provide another way of distinguishing between AGN and TDEs.  Near peak, the typical H$\alpha$ and \HeIIop\  FWHM of TDEs are both of order $10^4$ km s$^{-1}$ and often exceed $15,000$~km~s$^{-1}$ (Figure~\ref{fig:H_He_lines}).
Only a fraction of a percent of the
H$\alpha$ and H$\beta$ lines in
SDSS AGN (which may have problematic spectral decompositions) are as broad as in TDEs, i.e., $\gtrsim15,000$ 
(Figure~\ref{fig:agnfwhm}). This distinction may arise from the structure and dynamics of the BLR around SMBHs with certain masses \cite[][]{laor2003}.  Thus, line width is a reasonably good discriminator, but, depending on the overlap in the distributions for TDEs and AGN, may not be iron-clad.  

Whenever AGN spectra do have noticeable \HeIIop\ emission, the line profiles and widths are generally comparable to those of H$\beta$ (and thus also H$\alpha$). 
While accurate measurements for individual AGN are often challenging, given the weakness of the \heii\ feature and the fact that it is blended with several [Fe\,{\sc ii}] emission features, the resemblance between \heii\ and H$\beta$ can be seen in stacked spectra (e.g., Fig.~3 of \citealt{boroson2002}). 

The temporal evolution of the velocity line widths in response to changes in the photoionising continuum provides an even better way of distinguishing between TDEs and AGN.  In changing-look quasars, which are likely to contaminate TDE searches, the broad emission lines broaden as they weaken \citep{lamassa2015,runnoe}. This is the basis for reverberation mapping in AGN \citep[e.g.,][]{peterson1993} and the opposite of what is observed for TDEs \citep{Holoien:2016b}.

\paragraph{High He~\textsc{ii}/H$\alpha$ Ratio:\\}
\smallskip

The spectra of many UV/optical-bright TDEs have prominent
\HeIIop\ line emission. The line luminosity is typically of order $10^{41}$ erg s$^{-1}$ at its brightest, much stronger than H$\beta$ and comparable to (or even stronger than) H$\alpha$ (Figure~\ref{fig:H_He_lines}). Therefore, a \heii/H$\alpha$ flux ratio $\gtrsim 1$ in at least one spectral epoch is a hallmark of UV/optical-bright TDE candidates, as is \heii/H$\beta\sim1$.

\begin{figure*}	
\begin{center}
\includegraphics[width=0.6\textwidth]{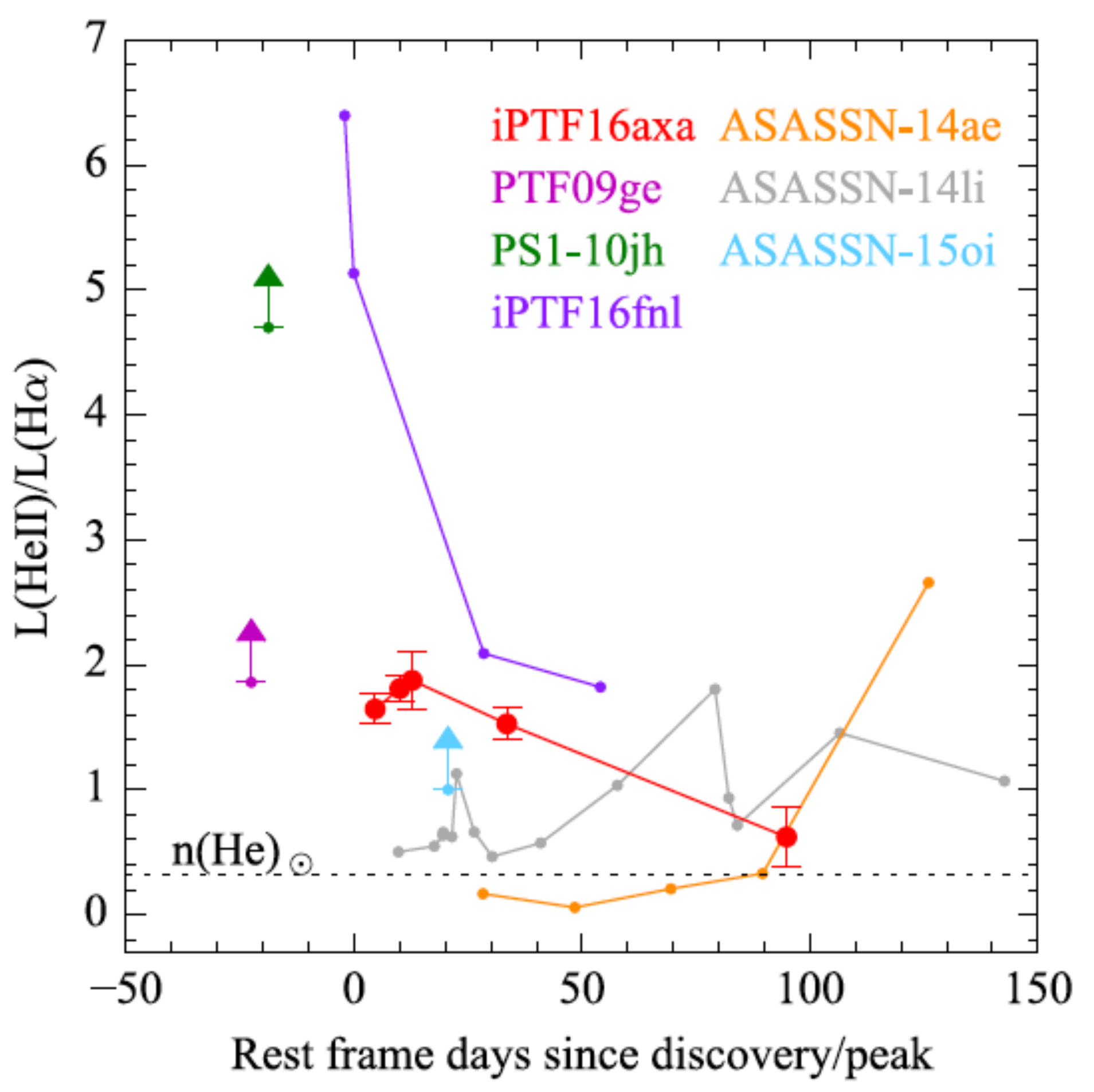}
\caption{The variation in the He~\textsc{ii}/H$\alpha$ flux ratio over time for a collection of TDEs. Observing this flux ratio $\gtrsim 1$ in at least one spectral epoch distinguishes optical/UV TDE candidates from most AGN. However, the ratio evolves with time, and so a single spectral epoch may miss the He~\textsc{ii} line if it appears at a different time.
Figure reproduced from \citet{Hung2017}. }
\label{fig:H_He_line_evolution}
\end{center}
\end{figure*}

There are potential exceptions. One possible case is TDE2 \citep[][the bottom spectrum in Figure~\ref{fig:H_He_lines}]{van-Velzen:2011a}, but its spectrum is low signal-to-noise. 
Another is PS1-11af \citep{Chornock:2014a}, although that event showed no emission lines at all in its spectrum.
Because the
He~\textsc{ii}/H$\alpha$ flux ratio is observed to evolve with time, we note that a single spectral epoch may miss the He~\textsc{ii} line if it appears at a different time (Figure~\ref{fig:H_He_line_evolution}).

In comparison, while AGN spectra do exhibit broad \HeIIop\ emission, it is typically weak compared to the Balmer lines.  The \heii/H$\alpha$ flux ratio in the SDSS quasar composite is $\sim 0.005$ \citep{VandenBerk2001}.  From the theoretical side, photoionisation modelling of the BLR in AGN also gives \heii/H$\alpha < 1$, although the goal of such work is usually to reproduce normal AGN spectra and not extreme outliers. 

Like other high-ionisation species and transitions, \heii\ comes from closer to the central engine than the Balmer lines \cite[e.g.,][]{Grier2013_2D_RM} and is extremely responsive to continuum changes in the AGN \cite[as in][]{KoristaGoad2004_BLR}, whereas  H$\alpha$ is the least responsive of the Balmer lines.  Thus, with a large (UV) flare in the AGN continuum (even if the SED shape does not change), it may be possible to boost the \heii/H$\alpha$ ratio temporarily.  
That said, while \citet{peterson1986}  describe a (moderate) flare in the AGN reverberation-mapping poster child, NGC~5548, that substantially boosts the \heii\ emission, it is always weaker than H$\beta$.  
As a caveat to the above, we must mention the measurement of the flux of the \HeIIop\ line can be affected by blending with the \NIIIbf\ line which can be excited by Bowen fluorescence \citep[e.g.,][]{Leloudas19,Onori2019,Nicholl2020}. Medium resolution spectroscopy may be helpful in deblending these two components.  

\subsubsection{UV Spectrum}

\begin{figure}
    \centering
    \includegraphics[width=0.75\textwidth]{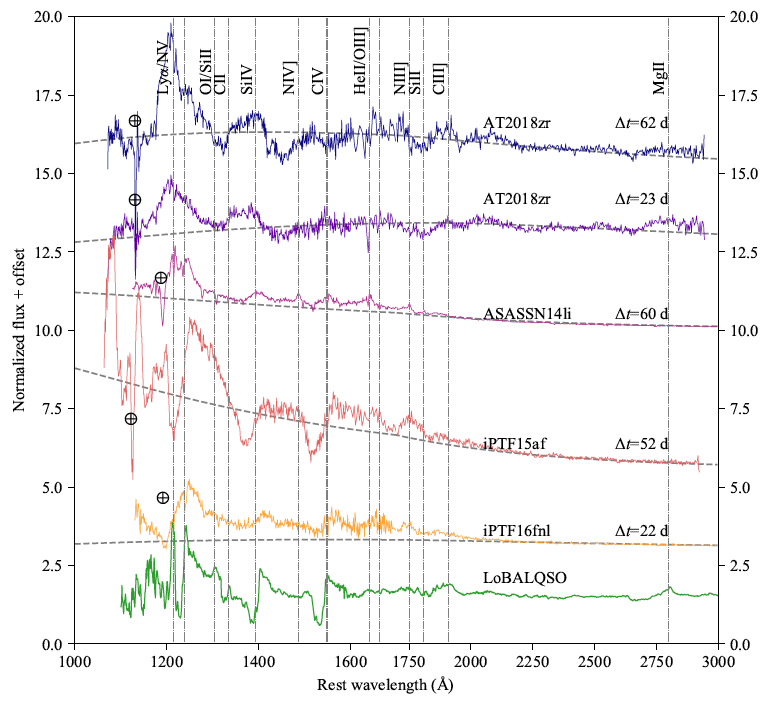}
    \caption{Ultraviolet spectra from four TDEs, including two epochs for AT2018zr, compared to a low-ionisation broad absorption line QSO (LoBALQSO) composite spectrum. At least three of the TDE spectra lack Mg\,\textsc{ii}\, $\lambda \lambda$ 2796, 2804 emission. 
    This figure appears in \citet{Hung2019} and is reproduced here with the permission of the American Astronomical Society.}
    \label{fig:uv_spectra}
\end{figure}

The UV spectra collected of TDEs so far (Figure \ref{fig:uv_spectra}) have revealed striking differences with respect to other AGN, although the sample is currently small. As of this writing, there are only 
two
TDE in quiescent galaxies with published UV spectra: 
iPTF15af \citep{Blagorodnova2019_iPTF15af} and AT2018zr \citep{Hung2019}, also known as PS18kh \citep{Holoien2018-2}. There is at least one more event with an unpublished UV spectrum, ASASSN-15oi (Alexander Dittmann et al., in prep.). Additionally, there is a near UV spectrum of PS16dtm \citep{Blanchard:2017a}, but, because that event took place in a narrow-line Seyfert I, we exclude it for the present purpose of distinguishing TDE from AGN. ASASSN-14li \citep{Cenko:2016a} and iPTF16fnl \citep{Brown2018} also have published UV spectra, but are not in quiescent host galaxies.

These systems generally lack certain low-ionisation emission lines that are common in most AGN, in particular Mg\,\textsc{ii}\, $\lambda \lambda$ 2796, 2804. The exception is AT2018zr, which displayed a broad emission feature consistent with Mg\,\textsc{ii}\, in five spectra taken between roughly 20 to 60 days after $r$-band peak. However, the equivalent width of Mg\,\textsc{ii}\, dropped steadily in time, and, by the final spectrum, the line had all but disappeared, while an absorption feature appeared at a blueshifted velocity consistent with the velocity (15,500 km s$^{-1}$) inferred from the Balmer lines in the optical spectrum \citep{Hung2019}. This rapid spectral variability distinguishes the event from most AGN in its own right.

Equally interesting is the general absence of C\,\textsc{iii]}\,$\lambda$ 1909 emission in the TDEs, a line seen in most AGN UV spectra. Adding to the differences is the general strength of N\,\textsc{iii]}\,$\lambda$ 1750 emission, a line which shows up prominently in only approximately 1\% of SDSS AGN. Here again AT2018zr is an exception, but only in the sense that its N\,\textsc{iii}\, does not show up clearly in emission, although it may be contributing to absorption at that wavelength. While the rare, so-called ``N-rich QSOs" do have have this emission line, they generally also have a strong C\,\textsc{iii]}\, line accompanying it, as well as Mg\,\textsc{ii}\,, which is not the case for TDEs \citep{Jiang2008,Cenko:2016a}.

There are broad absorption features in some UV TDE spectra, although these too distinguish themselves from broad absorption line quasars (BALQSO). TDEs with clear or tentative absorption include
iPTF16fnl, 
iPTF15af, and AT2018zr (PS18kh). In these cases, the absorption seems to correspond to the C\,\textsc{iv}\ $\lambda\lambda$1548,1551, Si\,\textsc{iv}\ $\lambda \lambda$1394, 1403, and N\,\textsc{v}\ $\lambda \lambda$ 1239,1243 lines.

The FWHM of these lines is roughly in the range 5000-10,000 km s$^{-1}$, quite similar to BALQSOs. However, in iPTF16fnl and 
iPTF15af, the centroid of the absorption lines is near enough to the line centre so that the absorption represents velocities from nearly 0 to 10,000 km s$^{-1}$, blending smoothly into the emission on the red side. In contrast, the centroid of the absorption lines in BALQSOs is often blueshifted by at least 10,000 km s$^{-1}$, and so the absorption is completely detached from the emission, with a broader wing on the higher velocity side \citep{Blagorodnova2019_iPTF15af}. AT~2018zr (PS~18kh) once again displays its own unique behaviour, with possible broad absorption centroid velocities of 15000 km s$^{-1}$; these absorption lines were also highly variable, becoming increasingly prominent with time in the five spectra taken over approximately 40 days. 

Of the TDEs with published UV spectra in quiescent hosts, both exhibit broad absorption lines at some point in time. This fraction is higher than that of BALQSOs. Blueshifted BALs in QSOs are thought to arise from fast-moving outflows. If the QSO/BALQSO dichotomy is largely due to viewing angle effects, then outflows in TDEs may subtend a larger solid angle than in AGN \citep{Hung2019}.

\subsubsection{X-ray Spectrum}

AGN activity could potentially mimic the X-ray emission arising from TDEs. In this section, we compare observations of X-ray emission from
AGN and TDE candidates, focusing on those that might help us distinguish between the two types of sources. The \xraychap\ has more information on the X-ray properties of TDEs.

\begin{figure}[t]
		\includegraphics[width=0.52\columnwidth]{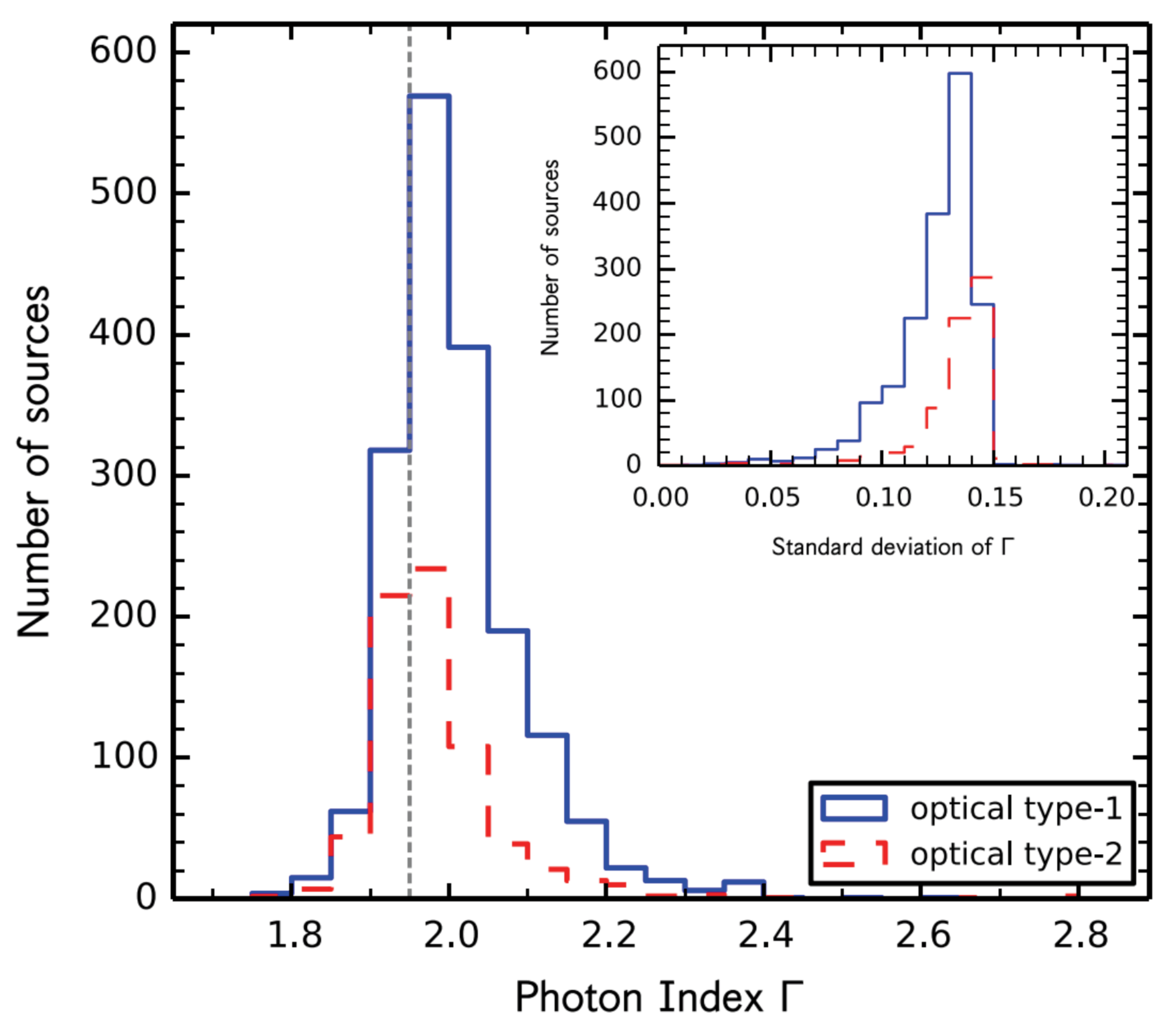}		\includegraphics[width=0.47\columnwidth]{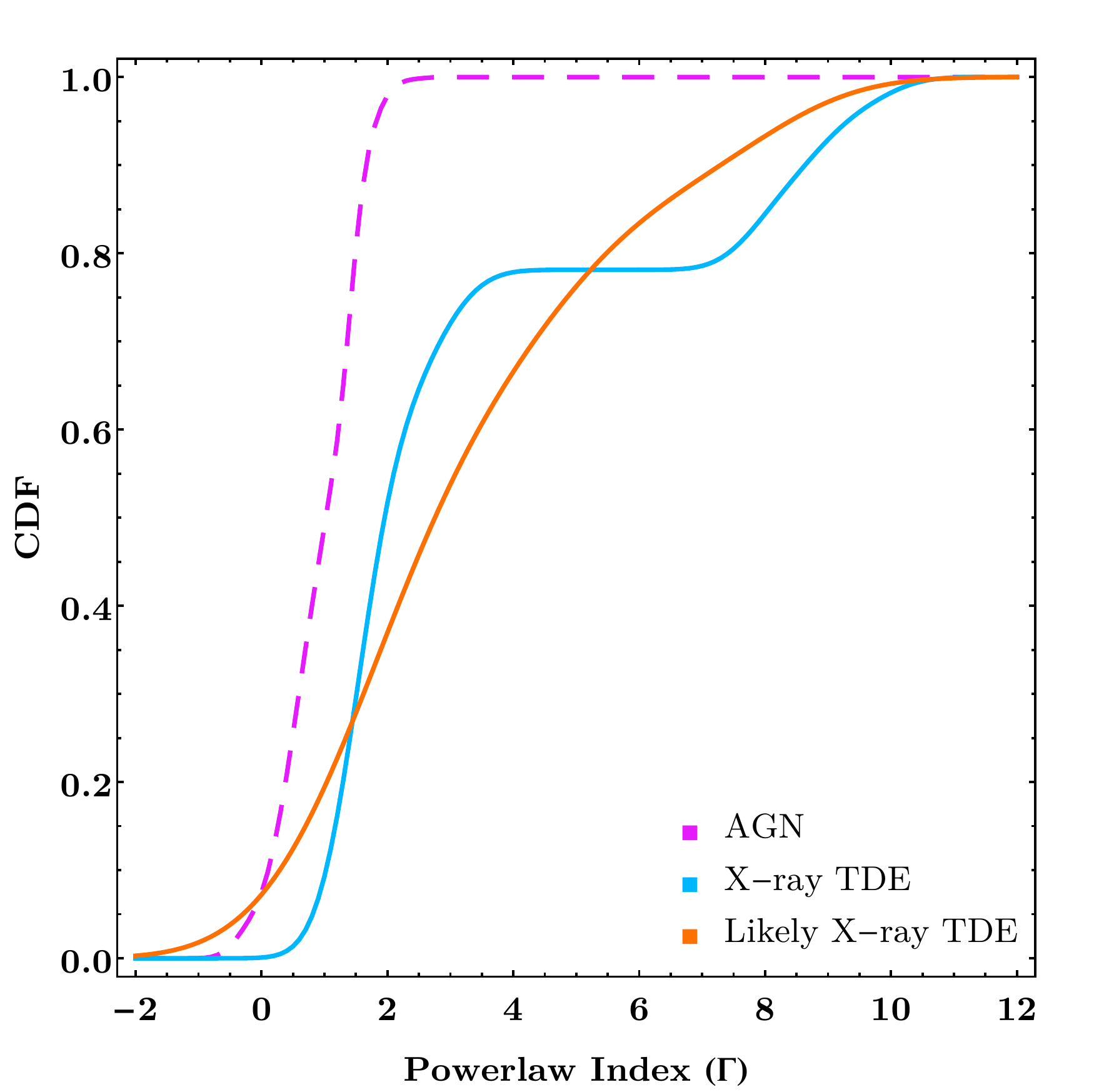}
		\caption{Comparison of power-law photon index $\Gamma$ for the X-ray spectra of AGN and TDEs.
		\emph{Left:} Distribution of photon indices of $\sim2500$ AGN observed with {\em XMM-Newton}. The AGN have non-thermal spectra with power law photon indices ranging from 1.7--2.4. Figure adapted from \citet{liu2016}.
		\emph{Right:} Cumulative $\Gamma$ distributions for events classified as X-ray TDEs (cyan curves) or likely X-ray TDEs (orange) by \citet{2017ApJ...838..149A} compared with AGN detected in the \emph{Chandra} Deep Field South  \cite[magenta;][]{tozzi2006,2011ApJS..195...10X}. Both TDE distributions from \citet{2017ApJ...838..149A}
		include jetted TDE candidates, which tend to have harder X-ray spectra than thermal TDEs and photon indices more like AGN.
		Observing a nuclear transient with steeper (softer) X-ray SED ($\Gamma\gtrsim3$) may strongly favour a TDE interpretation.
		} 
\label{Gammas}
\end{figure}

Thanks to its high-sensitivity and good sky coverage, the \emph{ROSAT X-ray Observatory} \citep{1982AdSpR...2..241T} discovered the first
TDE candidates. These nuclear transients had a peak X-ray luminosity of $L_x \sim 10^{44}$~erg~s$^{-1}$, were associated with galaxies that showed no evidence of (prior) AGN emission, produced light curves that decayed following a $t^{-5/3}$ power-law, and had X-ray spectra that were best described with a $\sim10^{5-6}$~K blackbody or with a very steep power-law index \citep[$\Gamma=3$-7, Figure \ref{Gammas};][]{1995A&A...299L...5G, 1995MNRAS.273L..47B, Bade:1996a, 1999A&A...343..775K, 1999A&A...350L..31G,Greiner:2000a}. 
Since \textit{ROSAT}, 
the capabilities 
of the \emph{Neil Gehrels Swift} Gamma-ray Burst Mission, the \emph{Chandra X-ray Observatory}, and \emph{XMM-Newton Space Observatory},
including increased effective area, spectral coverage, spatial resolution, and/or spectral resolution,
have dramatically changed our ability to characterise the detailed spectral evolution of TDE candidates, 
leading to discoveries including rapid variability and, in Swift J1644+57\footnote{See the \gammachap.}, possible jet formation \citep{Bloom:2011a,Burrows:2011a}.

\begin{figure*}	
\begin{center}
		\includegraphics[width=0.49\columnwidth]{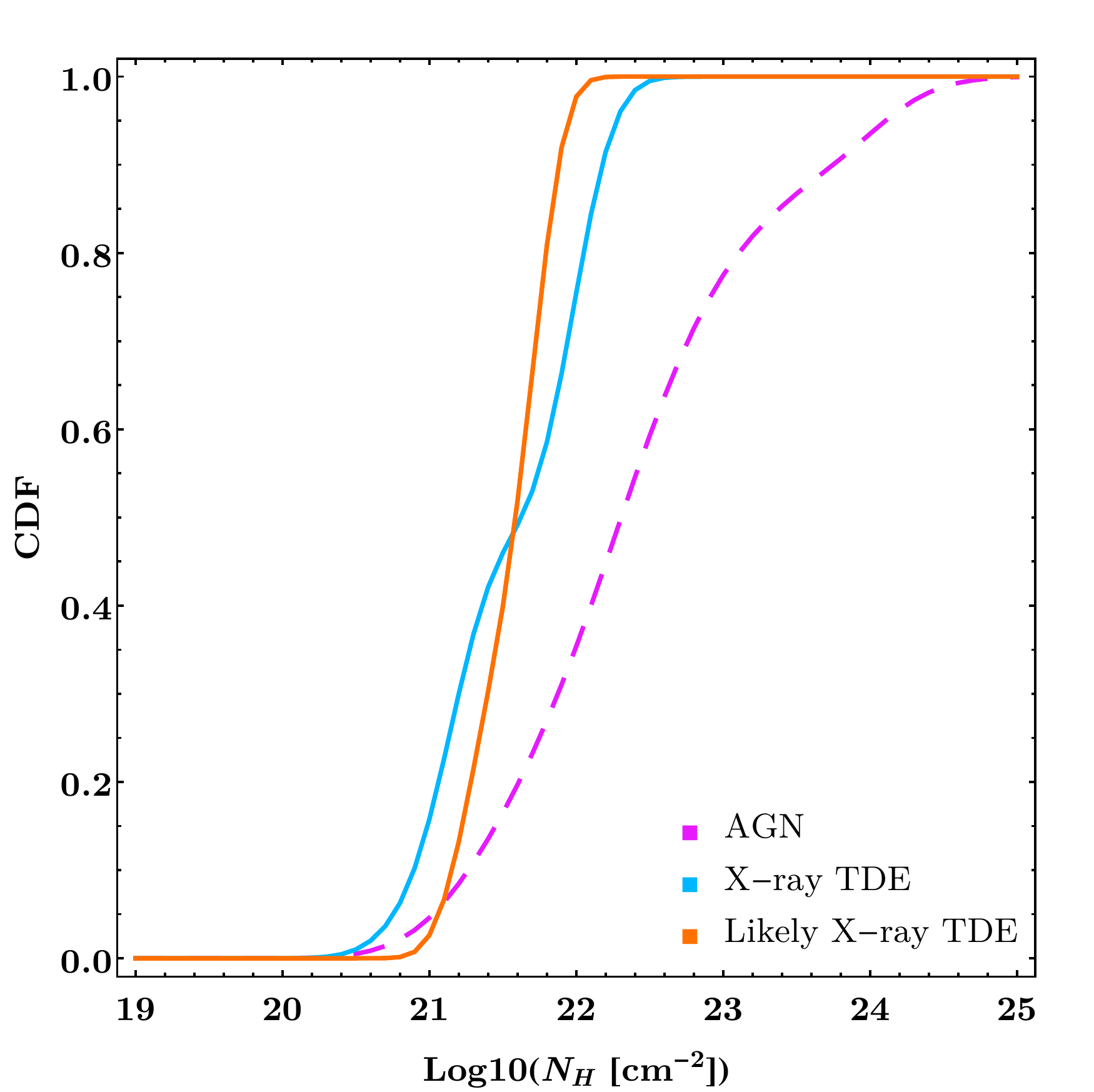}
\caption{Cumulative distribution function of line-of-sight hydrogen column density ($N_{\rm H}$) for TDEs classified as a X-ray TDEs (cyan curves) or likely X-ray TDEs (orange) by \citet{2017ApJ...838..149A}.
These TDEs are significantly less absorbed compared to AGN found at similar redshifts in the \emph{Chandra} Deep Field South  \cite[magenta;][]{tozzi2006,2011ApJS..195...10X}. It is not yet clear whether this difference is real or due to observational bias.
}
\label{NH}
\end{center}
\end{figure*}

AGN and TDEs are quite different in the nature and appearance of their broad-band X-ray spectra. 
In AGN, around 10\% of the bolometric luminosity is released as non-thermal X-ray continuum up to $\sim$100~keV,
as magnetic processes in and/or above the disk form a relativistic electron cloud that is partly cooled due to the inverse Comptonization of the thermal disk photons.
In the 2-10~keV band, this continuum is typically modelled as a simple power-law, whose index is determined by the temperature distribution of the electrons and by the number of electrons that each UV seed photon encounters.
In AGN, the average 2–10 keV spectral index is 1.9, with a spread from $\sim$1.7 to 2.4 at the most extreme (Figure~\ref{Gammas}; \citealt{liu2016}). 

In contrast, most thermal TDEs show X-ray emission that is dominated by a multicolor blackbody component with a temperature of 50-100 eV, e.g., ASASSN-14li \citep{Miller:2015b,Holoien:2016b,2017MNRAS.466.4904B}. In some cases, there is an additional weak hard X-ray tail, e.g., 
ASASSN-14li \citep{2018MNRAS.474.3593K} and XMMSL1 J074008.2-853927 \citep{Saxton2017_J0740}. For ASASSN-14li, \citet{2018MNRAS.474.3593K} suggest that the hard excess beyond the blackbody arises from additional inverse Compton scattering of disk photons by relativistic electrons in an X-ray corona. This additional Comptonization component appears to contribute more to the spectrum as the source evolves over time, either due to the accretion rate dropping or due to the delayed formation of an X-ray corona. Most recently, \citet{2020ApJ...897...80W} show that a ``slim disk'' accretion model adapted to SMBHs can successfully describe the multi-epoch X-ray spectra of ASASSN-14li.

Another observed distinction is that TDEs are significantly less absorbed compared to AGN found at similar redshifts (Figure \ref{NH}). We note, however, that this difference may arise from observational bias. A column density of 10$^{22}$ cm$^{-2}$ would completely absorb the emission of soft TDEs with a 50 eV thermal spectrum and hinder their detection.

\begin{figure*}	
\begin{center}
\includegraphics[width=0.6\textwidth]{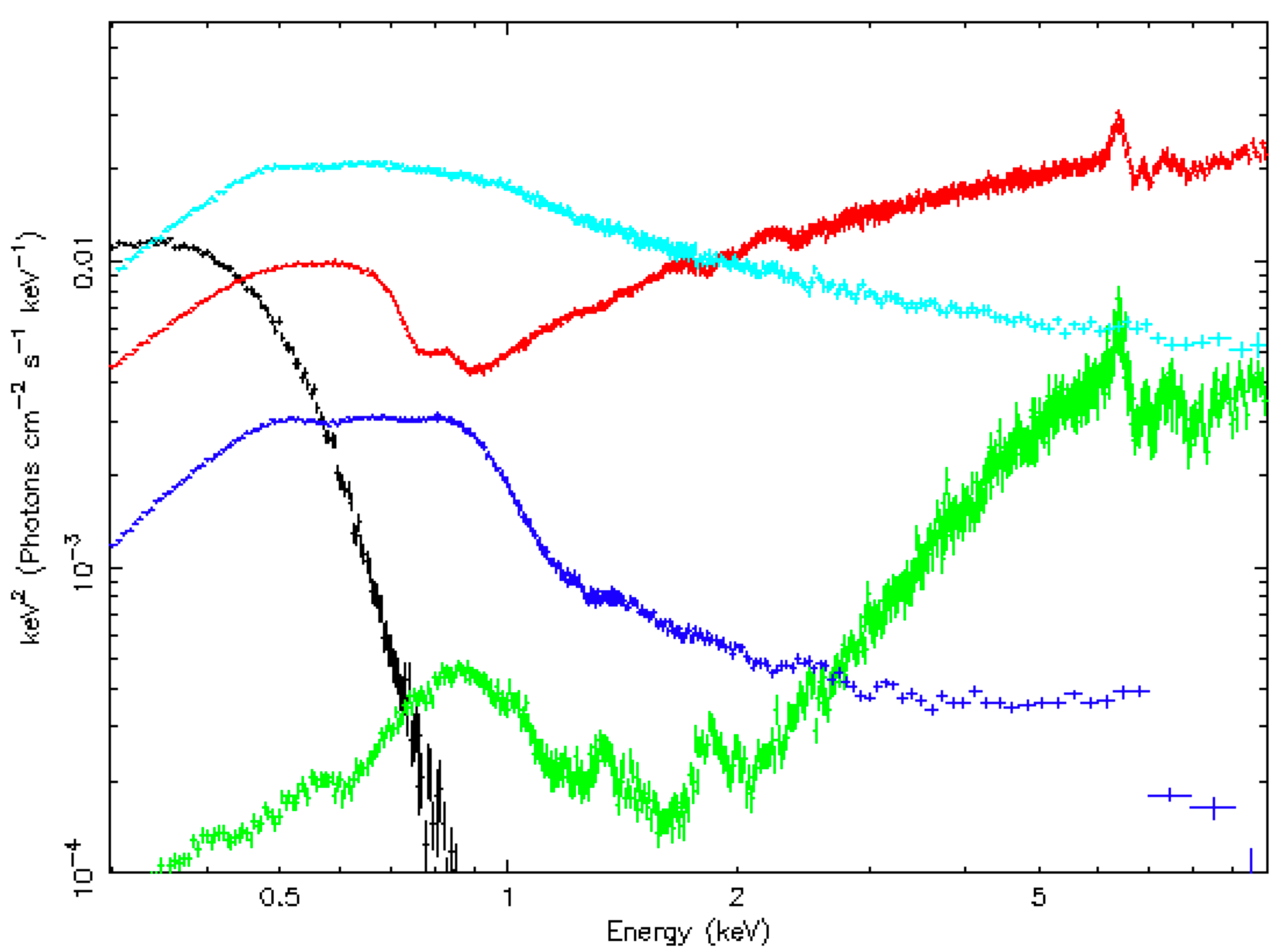}
\caption{Comparison between the X-ray spectra of the well-known thermal TDE ASASSN-14li (black) and AGN of different types, including MCG-6-30-15 (red), NGC~1365 (green), 1H0707-495 (blue), and Ark~454 (cyan). 
Even Eddington-limited Narrow-line Seyfert 1 AGN (in blue and cyan) do not show as steep an X-ray spectrum as thermal TDEs. Figure from E. Kara.}
\label{compare_Xray_spec}
\end{center}
\end{figure*}

The thermal-dominated X-ray spectra of TDEs such as ASASSN-14li differ from those of a wide range of AGN, from Type 1 (unabsorbed) to Type 1.9 (absorbed), all of which have non-thermal emission (Figure~\ref{compare_Xray_spec}). 
Two Narrow-Line Seyfert 1 AGN, 1H0707-495 and Ark 564, have been proposed as Eddington-limited or even super-Eddington AGN. Even these extreme accretion AGN show non-thermal X-ray emission and are distinct from ASASSN-14li. 

While thermal TDEs have softer X-ray spectra than AGN \citep{Lin2011,2017ApJ...838..149A}, the seemingly separate class of jetted TDEs, such as Swift J1644+57, is characterised by strong X-ray emission up to $\sim 100$~keV and an AGN-like photon index \citep[][see also Section 5]{Bloom:2011a, Burrows:2011a}. Complicating the picture further are observations showing that some thermal TDEs spectrally harden as they evolve \cite[e.g., ASASSN-14li;][]{2018MNRAS.474.3593K}, while others show little variation.
For the time being, we conclude only that a nuclear transient with a steeper (softer) X-ray SED ($\Gamma\gtrsim3$) is more likely to be a TDE than an AGN.

\subsubsection{X-ray Variability}

The variable X-ray emission observed in persistent AGN spans timescales from seconds to months and years, with an
amplitude often much larger than at optical wavelengths \cite[for the same source; e.g.,][and references therein]{Lanzuisi2014_Xvar}.
In contrast, the X-ray (and optical/UV) light curves of TDEs show a steady, months-long structure.
Hence, measuring a transient's X-ray light curve with a short cadence and over years, and comparing it to the UV/optical light curve, may prove
an important tool in identifying TDEs by
excluding those AGN that vary on much shorter or longer timescales.

It is not yet clear how strongly
the observed decay rate of the X-ray light curve
discriminates between TDE and variable AGN.
The self-regulating nature of the SMBH accretion in AGN can produce an X-ray light curve that declines as $t^{-(1.5-2.0)}$ \citep[e.g.,][]{2009ApJ...698.1550H}.
While a power-law of $t^{-5/3}$ has been associated with some TDE optical/UV light curves and the first X-ray light curves, 
the observed temporal evolution of soft X-ray emission from current X-ray and likely X-ray TDE candidates  \citep{2017ApJ...838..149A} follows a wide variety of power-laws, consistent with fall-back, accretion, and disk emission \citep[e.g.,][]{1989ApJ...346L..13E,1989Natur.340..595P,1990ApJ...351...38C, 2009ApJ...700.1047C, 2011ApJ...742...32C,Lodato:2011a,Guillochon:2013a}, with the majority shallower than $t^{-5/3}$. \citet{Guillochon:2013a} suggest that such shallow declines arise when
TDEs are viscously delayed (i.e., the time it takes for material to accrete is slow).


Many AGN of extreme X-ray variability have been discovered thanks to the photon-counting nature of X-ray facilities, as well as the long integration times used for some sources and for certain extragalactic fields. These AGN remain poorly understood, due to the frequent lack of simultaneous data at other wavelengths. The X-ray spectra of  “X-ray changing-look” AGN \citep{Matt2003_CLAGN_Xrays,2012AdAst2012E..17B,Ricci2016_IC751_CLAGN} have been well modeled by occulting clouds transiting into and out of the line-of-sight towards the AGN, alternately masking and uncovering the central engine. However, the X-ray spectra of some optical changing-look AGN, with their weakening or strengthening of broad Balmer lines, cannot be ascribed to variable extinction. When observed in the dim state, the X-ray spectra do not show features of obscuration that extinguishes and reprocesses the X-ray emission, but rather are consistent with a model where the the intrinsic X-ray emission diminished significantly \citep[e.g.,][]{lamassa2015,lamassa2017,husemann2016}. The processes responsible for driving the X-ray variability are unknown, but there may be a link between the Eddington ratio and spectral state of the AGN due to mechanisms that operate at the scales of the inner accretion disk, similar to the spectral state transitions observed in stellar mass black hole binaries \citep[e.g.,][]{Noda2018, Ruan2019}.

TDEs tend to show relatively little variation in column density with time \citep{2017ApJ...838..149A}; after an initial drop, $N_H$ remains roughly constant over at least several hundred days in both ASASSN-14li and -15oi \citep{2020ApJ...897...80W}.
Thus, observing an evolving column density is a possible way of distinguishing some AGN from TDEs. We note, however, that the X-ray TDEs discovered to date tend to have lower $N_H$ than AGN (Figure \ref{NH}), which may limit the measured $N_H$ variation.
For cases where declining absorption can be excluded as the cause of X-ray brightening,
the distinction between TDE and X-ray variable AGN is less apparent. For instance, in the Seyfert 1.9 galaxy IC 3599, TDEs were alternately invoked \citep{Campana:2015a} and ruled out \citep{Grupe:2015a} to explain the observed, recurring X-ray flares.

\subsection{Differentiating TDEs from Flaring AGN} \label{sec:Flaring_AGN}

\subsubsection{Accretion due to Disk Instabilities}

In contrast with steady-state AGN disk accretion, accretion events arising from AGN disk instabilities might be relatively quick and deplete only the innermost AGN disk, a scale comparable to that expected for TDEs.
Indeed, fast and coherent instabilities in the innermost parts of the disk seem to be required to explain significant variability events in accreting SMBHs, as this is where most of the (continuum) radiation is produced 
(see, e.g., \citealt{Lyubarskii1997}, \citealt{King2004}, and the discussion in \citealt{Cannizzaro2020_Gaia16aax}).
Such events might even produce higher accretion rates and luminosities than during the steady-state and/or quiescent stages. In other words, AGN flares might share many similarities with TDEs.

Thermal, viscous, gravitational, and radiation-pressure instabilities have all been suggested to occur in AGN disks (e.g., \citealt{Jan+02,Jia+13} and references therein). Below is a brief review of the main aspects of such instabilities and their implications. We generally follow the discussion in \citet{Saxton:2015a}. 

Accretion disks may attain a limit-cycle behaviour, which can be generally divided into three phases: in the 1) \emph{quiescent phase}, material initially accumulates at a slow rate and fills the inner region of the disk. The disk structure and properties then slowly evolve until the disk becomes unstable, leading to the 2) \emph{rise and outburst phase}, where the instability typically leads to a runaway heating, increasing the local viscosity, scale height of the disk, and the local accretion rate \citep{Cannizzo96}. This process eventually changes the accretion rate and produces a rapid depletion of the unstable region, whose material then accretes onto the SMBH. Such a fast accretion episode leads to a flaring of the SMBH. As the inner disk depletes, when the matter is accreted into the black hole
faster than it is replenished, the accretion flare goes through the 3) \emph{decay phase}, leading to a new quiescent phase and the next accumulation cycle. 

The disk-instability model thus predicts \emph{repeated} flares. The possibility of observing multiple flares depends on the typical duty-cycle timescale. 
For an unstable disk model to explain the known TDE candidates, the timescale for the limit-cycle must be sufficiently long, given the non-detection of repeated TDEs to date.
\footnote{The (in)ability to detect such repeated flares naturally also depends on their amplitudes, which are not necessarily as high as the flare that led to the identification of the TDE candidate.
} 

The overall depletion of the inner unstable disk is likely to occur on the timescale $\tau_{\rm{dep}}$ for material to viscously accrete from the truncation radius, the outermost part of the unstable region down to innermost stable circular orbit (ISCO). The initial rise should be fast, as the accretion begins from the innermost region, and then slower, up to the timescales for the material in the outermost truncation radius region to accrete down to the SMBH. Given the mass enclosed in this region and the accretion rate, one can estimate a typical timescale for the overall rise and fall:

\begin{equation}
   \tau_{\rm{dep}} = M_{\rm{inner}} / \dot{M}.
\end{equation}
The enclosed mass of the inner disk is given by 

\begin{equation}
   M_{\rm{inner}}= \int_{R_{0}}^{R_{\rm{trunc}}}\rho(r)\,2\pi r\,H(r)\,dr,
\end{equation}
where $R_{0}$ is the radius of the ISCO, $R_{\rm{trunc}}$ is the truncation radius, $\rho(r)$ is the disk density, and $H(r)$ is its height.
For a Shakura-Sunyaev thin disk \citep{shakura73} and typical values, expressed in gravitational radii, the enclosed mass is
\begin{equation}
    M_{\rm{inner}} = 6 \times 10^{-4}\alpha^{-8/10} M_{6}^{11/5} \dot{M}_{\rm{Edd}}^{-3/10} \left[\left(\frac{R_{\rm{trunc}}}{R_{\rm{g}}}\right)^{5/4} -  \left(\frac{R_{0}}{R_{\rm{g}}}\right)^{5/4}\right] 
    \,{M_{\odot}}\, ,
\end{equation}
where $\alpha$ is the viscosity parameter, and the accretion rate is given in units of the Eddington-limited accretion rate,
$\dot{M}_{\rm Edd}\simeq1.4\times 10^{24} M_{6}$ g s$^{-1}$,
for a SMBH mass $M_{6}$ in units of $10^{6}\,M_{\odot}$  \citep{Saxton:2015a}. 
The depletion time  $\tau_{\rm dep}$ (with typical truncation radius of a few tens of gravitational radii) is then
\begin{equation}
    \tau_{\rm dep} \sim 0.33\,{\alpha_{uns}}^{-8/10}\,M_{6}^{6/5}\,\dot{M}_{\rm{Edd}}^{-3/10}
    \left[\left(\frac{R_{\rm{trunc}}}{R_{\rm{g}}}\right)^{5/4} - \left(\frac{R_{0}}{R_{\rm{g}}}\right)^{5/4}\right] \,\mathrm{months}.
\end{equation}
In other words, $\tau_{\rm dep}$ is equivalent to the viscous
timescale of a thin disk at the truncation radius, and $\alpha_{uns}$ is the viscosity in the unstable region, rather than the typical viscosity operating during the regular accretion phases. 

There are many uncertainties both in understanding the disk instability process and in the estimates of the typical timescales and accretion rates, making this approach simplified at best. Much depends on
the choice of 
$R_{trunc}$ and on whether these instabilities occur there.
Nevertheless, the reasoning here suggests that the timescales for the rise and fall of such flares 
for low mass SMBHs (up to $\sim10^7 \,M_{\odot}$)
are on the order of weeks to months, while the timescales for more massive SMBHs are probably too long for the flares to be classified as fast transients.
In other words, disk-instability flarings potentially masquerade as TDEs of MS stars only for lower mass SMBHs. 
In addition, for SMBH masses larger than $\sim10^8$ M$_{\odot}$, the ionisation region is also Toomre unstable (Q$<$1, self-gravity).

Interestingly, the timescales of both TDEs and disk-instability flares depend on the SMBH mass, but through different scalings. This difference could potentially be used to distinguish between the models. Nevertheless, better comprehension of both the TDE process (including TDE debris disk circularisation) and disk instabilities that lead to flares is needed before reliable predictions and comparisons can be made. 

As noted by \citet{Saxton:2015a},
the overall timescale for the duty-cycle, and hence for repeating flares, is determined by the filling or viscous time at the truncation radius.
The viscosity in the stable region beyond the truncation radius differs from that in the inner region during the depletion, and so refilling takes far longer than the depletion and flaring timescale, i.e., decades or more for  $M_{\rm{BH}}\sim10^{6} M_{\odot}$. 
Repeated flares on such a timescale have been observed in the the Sy 1.9 galaxy IC~3599, 20 years apart \citep{Grupe:2015a}.

\begin{figure}[t]
	\begin{center}
		\includegraphics[width=0.49\columnwidth]{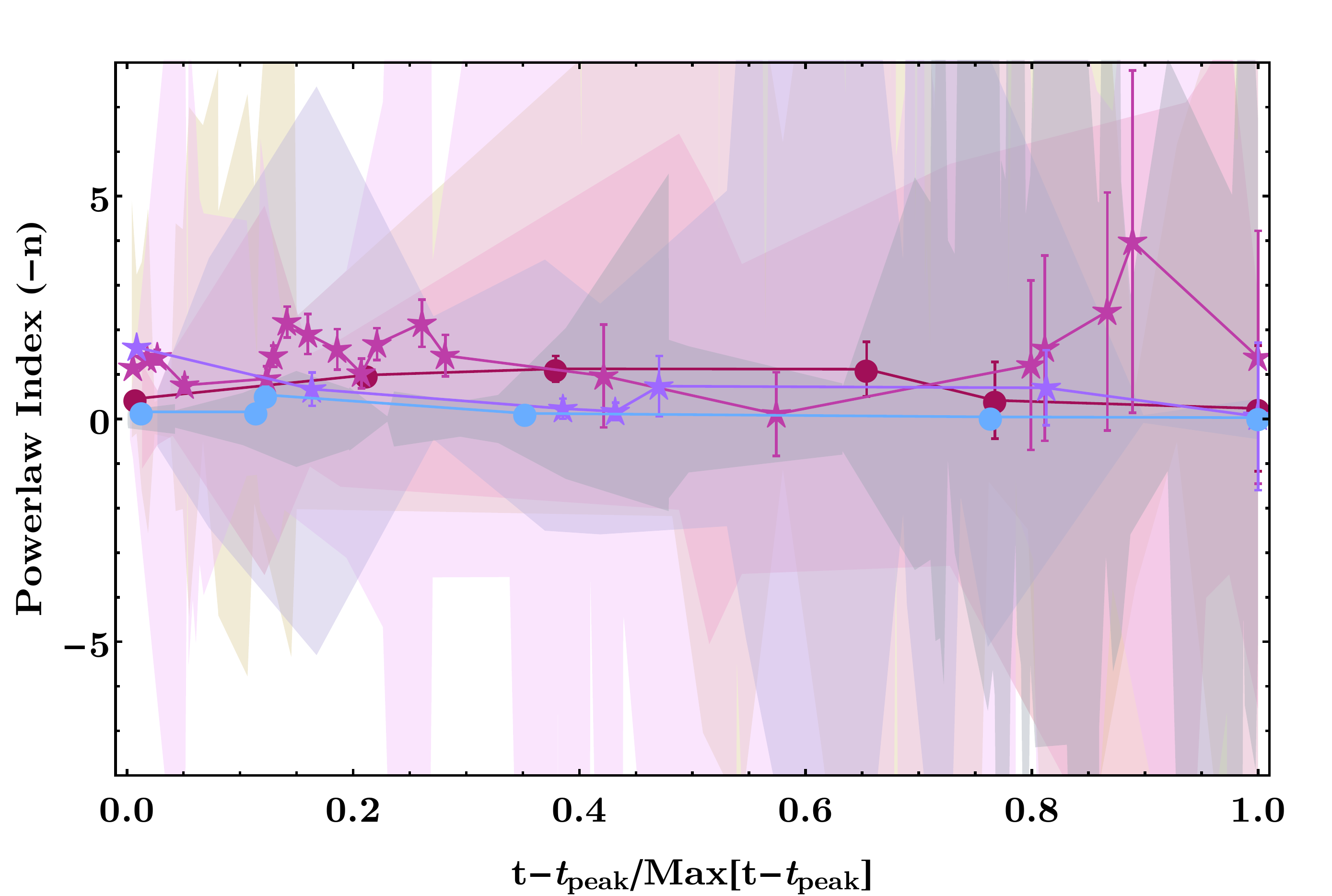}
		\includegraphics[width=0.49\columnwidth]{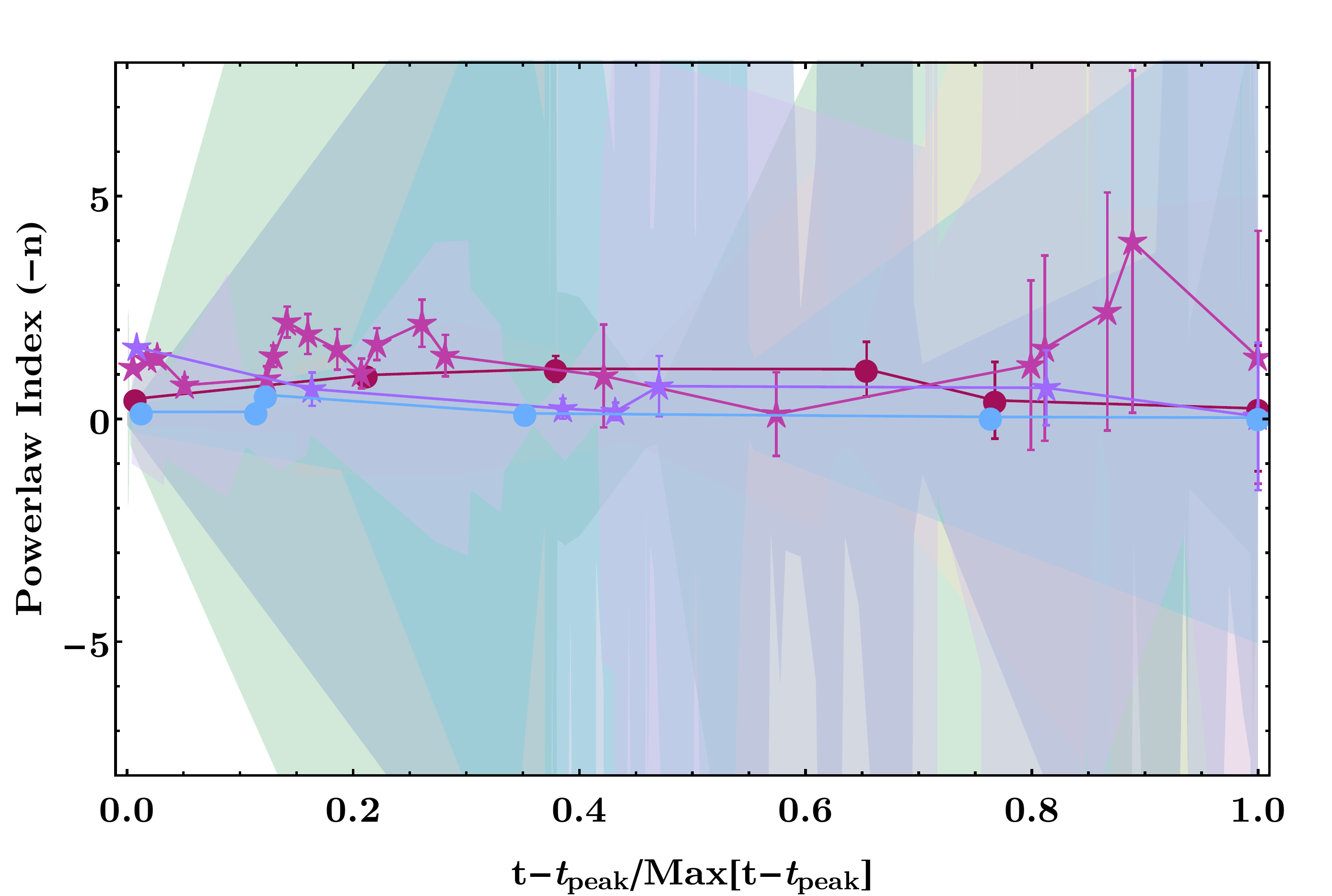}
		\includegraphics[width=0.6\columnwidth]{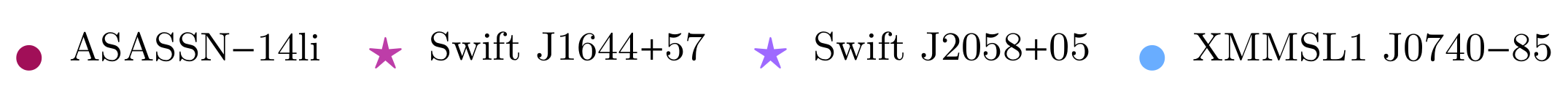}\\
		\includegraphics[width=0.5\columnwidth]{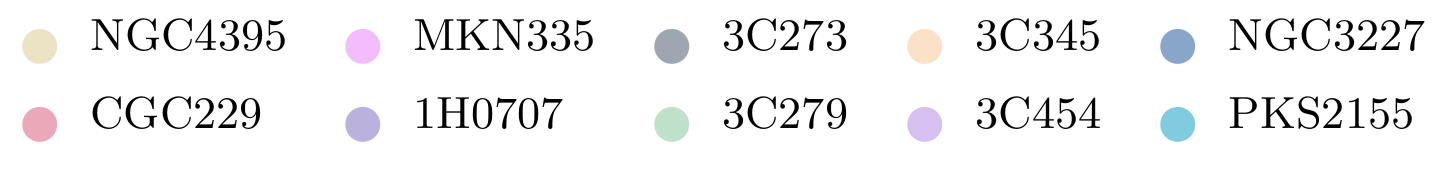}
		\caption{Comparing the smoothness of X-ray light curve decay for TDEs and AGN flares. \citet{2018ApJ...852...37A} determine the best fit power-law index ($-n$) for TDE and AGN samples as the time of peak goes to infinity, assuming 
		$L \propto (t - t_{peak})^{-n}$.  Here we plot $n$ as a function of $t - t_{peak}$, where $t_{peak}$ is the time in which the peak luminosity is detected. The sources decay over different timescales, so $t - t_{peak}$ is normalised by its maximum value for each source. The derived power-law indices for AGN vary significantly over short time scales, ranging between $n \sim -10$ and $+15$. Different colour shaded bands represent the full range of these indices and their uncertainties. In contrast, the decay of a TDE flare is best fit with a power-law index between $n = 0$ and $\sim -2$; for most of the decay, the power-law index changes little, especially for the two non-jetted TDEs ASASSN-14li and XMMSL1 J0740-85. Figure adapted from  \citet{2018ApJ...852...37A}.\label{hvavstde2}}
	\end{center}
\end{figure}

Due to the uncertainties in the duty cycles of AGN, it is not always feasible to rule out
recurring flare emission due to an AGN
when seeking to confirm a TDE. Another difficulty is that, during
an AGN flare or TDE, the change in source luminosity, relative to the pre-event upper-limits, is similar
\citep{2018ApJ...852...37A}. However, \citet{2018ApJ...852...37A} find that TDE X-ray light curves decay much more coherently, even monotonically, while the rate of AGN decay varies widely with time (Figure \ref{hvavstde2}). 
Less than 4\% of the coherent decay behaviour seen in their TDE sample could arise from sources like those in their AGN sample, suggesting that observing smooth decay can help to distinguish TDEs from AGN flares.

Galaxies hosting \emph{known}
AGN are typically excluded from TDE candidate selection, given the confusion arising from the expected AGN variability and flaring.
However,
galaxies hosting undetected, sub-luminous (``starved") AGN may present as quiescent. Given the low accretion rates in starved AGN \citep{Saxton:2018}, their disk properties and variability could differ from those of persistent AGN, and their occasional flares might masquerade as TDEs.
Gas-poor hosts are less likely to enable gas inflows to the nucleus, so flarings in starved AGN may favour ``gas-intermediate" hosts, whose histories include higher rates of star formation in the past.

\cite{Saxton:2018} suggests that the apparent preference of TDE candidates for ``post-starburst" galaxies (see the \hostchap) naturally arises from the reasons above and that many of these events are in fact AGN flares. One prediction of the 
disk instability scenario is therefore
that deeper observations of TDE candidate hosts will reveal weak AGN, either now or in the recent past, at higher rates than in other, comparably massive galaxies. However, at least for known low luminosity AGN (if defined by LINER emission), it is unlikely that UV/optical-bright TDE candidates arise from tail end of normal Type 2 AGN variability (Section \ref{host_galaxy_prop}).

\subsubsection{Flaring of Known AGN}
\label{ref:flares_in_known_AGN}

\begin{figure*}
    \centering
\includegraphics[trim={0 0 8.7cm 1.75cm},clip,height=0.375\textwidth]{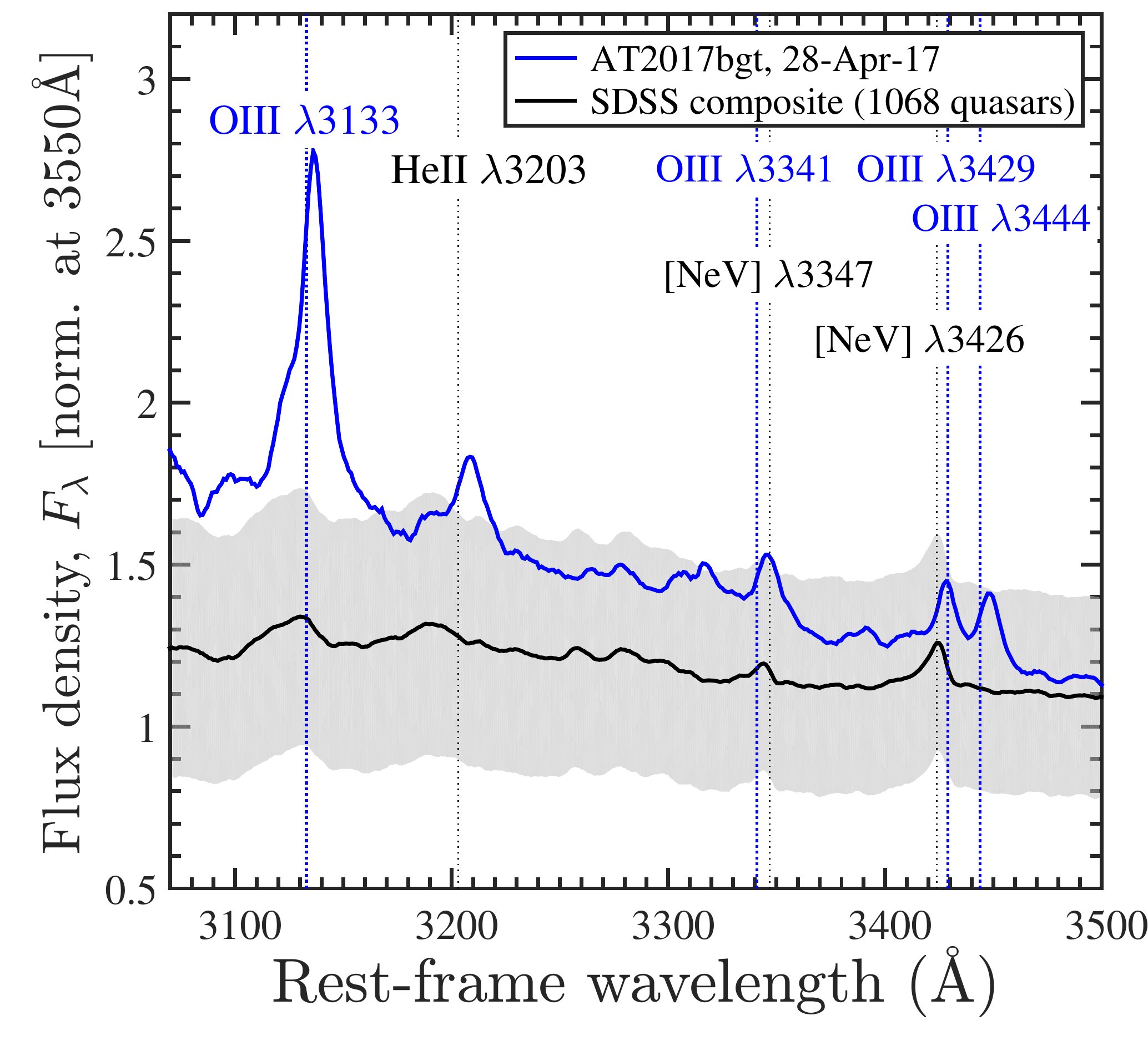}
\hfill
\includegraphics[trim={0 0 0 1.75cm},clip,height=0.375\textwidth]{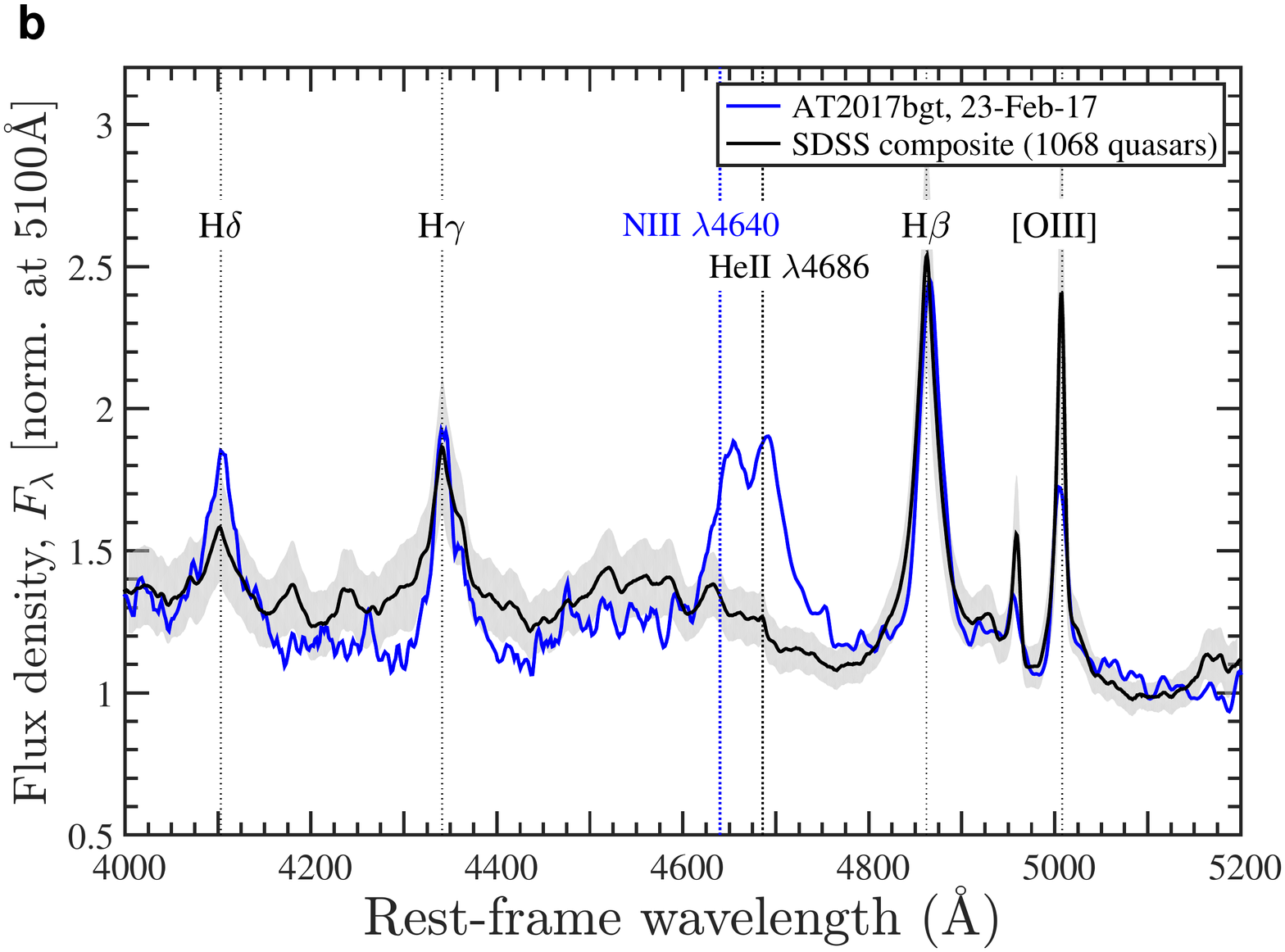}
\caption{Optical spectra of AT~2017bgt and unobscured AGN (adopted from \citealt{Trakhtenbrot2019_AT2017bgt}).
Two spectra of AT~2017bgt (blue), taken at different epochs within about two months of discovery, are compared to a composite of more than a thousand SDSS broad-line AGN (quasars) with similar hydrogen emission line widths. 
The broad Balmer lines and narrow forbidden [O\,{\sc iii}]\,$\lambda\lambda4959,5007$ lines of AT~2017bgt (and other events in this new class) are similar to those in the AGN. On the other hand, the prominent double-peaked emission feature near 4680\AA\ (\emph{right panel}), 
the prominent \OIIIbf\ and He\,{\sc ii} $\lambda3203$ lines, 
and the weaker O\,{\sc iii} $\lambda\lambda3341,3429,3444$ lines (\emph{left}), are not seen in the AGN.
These features arise from Bowen fluorescence, indicating an atypically strong source of high-energy (X-ray/EUV) radiation that produces intense \heii\ emission, which in turn drives the O\,{\sc iii} and N\,{\sc iii} emission lines through multiple scatterings and excitations in an optically thick medium.}
\label{fig:AT2017bgt_comp_SDSS}
\end{figure*}

\citet{Trakhtenbrot2019_AT2017bgt} identify a new class of flares from accreting SMBHs, which may be of particular importance to TDE classification.
The light curves, which are exemplified by AT~2017bgt, the brightest and best-studied flare, exhibit a significant increase in UV/optical emission, followed by a long, slow decline, on timescales of a year (or more). The optical spectra show both narrow and broad emission lines, most of which
resemble those of AGN, particularly NLSy1.
Most importantly, AT~2017bgt-like events have a prominent double-peaked emission feature near 4680 \AA, which is composed of the \HeIIop\ and \NIIIbf\ emission lines, and several other strong O~\textsc{iii} transitions, all with widths similar to the broad Balmer lines. These lines, driven by Bowen fluorescence (BF), are not seen in normal AGN (Figure \ref{fig:AT2017bgt_comp_SDSS}), despite specific predictions \citep{Netzer1985_HeII}.
Their existence in these UV-bright transients indicates that the BF process in dense gas near accreting SMBHs requires an exceptionally strong incident UV continuum. 

\begin{figure}
\centering
\includegraphics[width=0.65\textwidth]{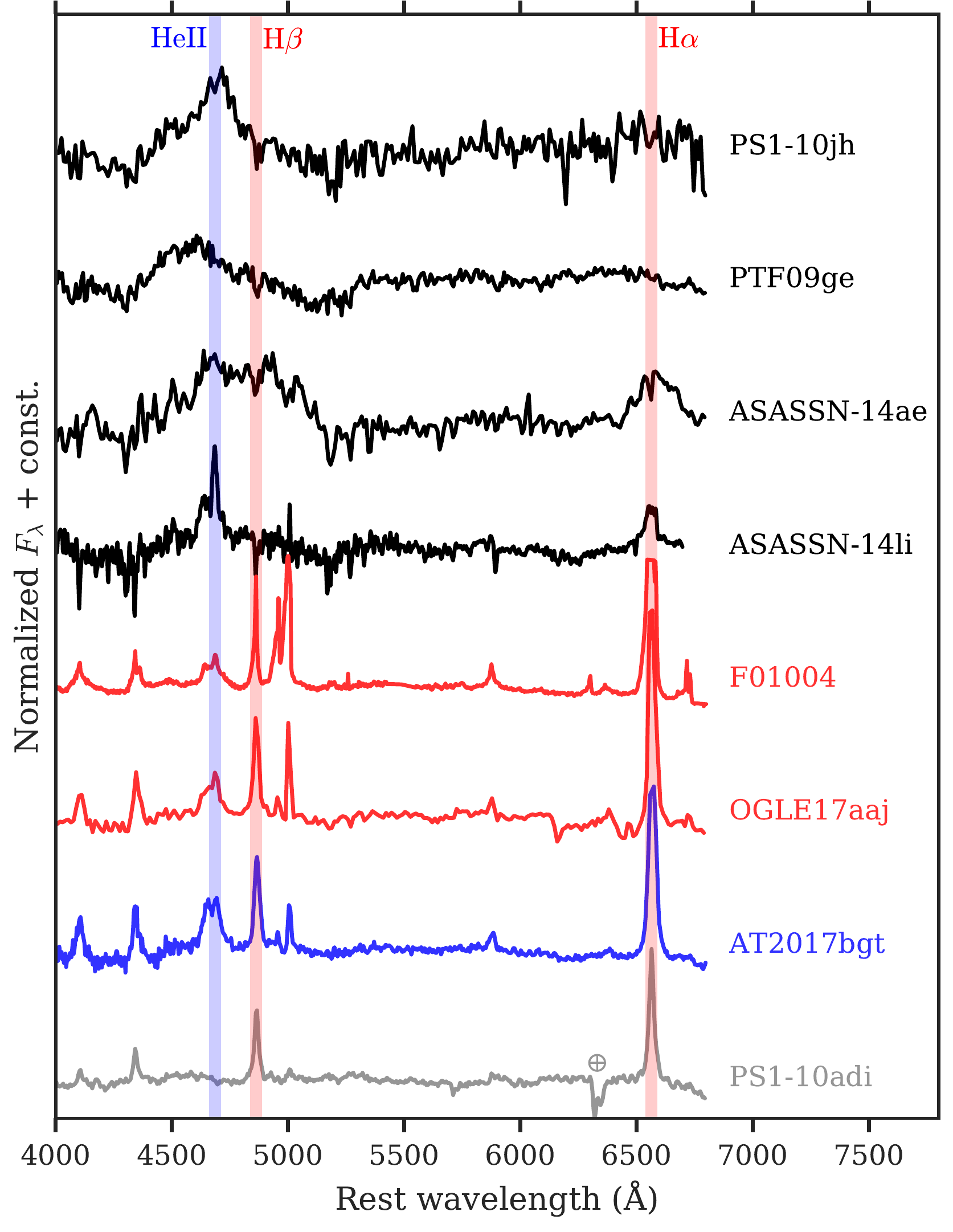}
\caption{Broad emission features near \HeIIop\ in AT~2017bgt, and similar objects, compared to other nuclear transients (adapted from \citealt{Trakhtenbrot2019_AT2017bgt}).
The spectra of AT~2017bgt (blue) and the events in F01004-2237 \citep{Tadhunter:2017a} and OGLE17aaj \citep[][both in red]{Gromadzki2019_OGLE17aaj} represent a new class of nuclear transients.
Also shown are the spectra of four TDEs \cite[from][in black]{Gezari:2012a,Arcavi:2014a,Holoien:2014a,Holoien:2016b} and of the luminous, slowly-evolving transient PS1-10adi \cite[][in grey]{Kankare2017_PS10adi}.
All spectra are continuum-subtracted.
The feature near 4860 \AA\ in the AT~2017bgt-like events, which originates from \HeIIop\ and the Bowen fluorescence \NIIIbf\ transitions, is significantly narrower than what is typically seen in most TDEs.}
\label{fig:AT2017bgt_spec_comp_TDEs}
\end{figure}

While this new class of UV-bright flares from SMBHs 
was identified from only three events, including the transient in the ultra-luminous infrared galaxy F01004-2237 \citep{Tadhunter:2017a} and the OGLE17aaj event \citep{Gromadzki2019_OGLE17aaj},
on-going transient surveys should detect additional events of this kind.
Such events are ``TDE impostors,'' at least initially, as their early optical spectra would show a strong, broad feature near 4680 \AA, which could be interpreted as \HeIIop, a common emission line in optical TDE candidates.
Indeed, one of these events was originally classified as a TDE \citep{Tadhunter:2017a}.
However, this feature, which is composed of two separate emission lines, 
is both narrower and weaker than the \HeIIop\ 
emission observed
in TDE candidates to date, i.e., with smaller FWHM {\it and} flux ratios relative to Balmer lines
(see Figure \ref{fig:AT2017bgt_spec_comp_TDEs}).

Other recently discovered transients, some of which are strong TDE candidates, exhibit BF features in their optical spectra.
TDE examples include 
iPTF15af \citep{Blagorodnova2019_iPTF15af},
iPTF16fnl \citep{Onori2019}, ASASSN-18pg/AT2018dyb \citep{Leloudas19},
and ASASSN-14li, which has some evidence for such features \citep{Holoien:2016b}.
Newer work shows that, 
in a flux-limited sample, TDEs with Bowen lines (or ``TDE-Bs") are as common as TDEs with only broad hydrogen lines; in a volume-limited sample, TDE-B's are the \emph{most} prevalent of the three TDE classes considered by \citet{vVelzen2020}.

Thus, the BF mechanism appears to frequent a range of nuclear, UV/optical-bright transients, all of which are related to enhanced accretion onto a SMBH. 
In some cases, the optical light curve, the peak luminosity, and the historical lack of SMBH activity, strongly favour the TDE interpretation. 
In other cases, where there is robust evidence for a pre-existing AGN (e.g., AT~2017bgt), we must be more prudent. While the nature of AT~2017bgt-like events remains ambiguous, the slow evolution of their light curves also argues against a ``simple" TDE origin. One intriguing possibility is that these events arise from a tidal disruption stream colliding and interacting with a pre-existing AGN accretion disk. 
As noted by \citet{Chan19}, testing this scenario requires simulations that fully explore the geometry and orientation of the tidal stream relative to the disk, the density and velocity structure of the stream, and the properties of the unperturbed disk, as well as that predict the radiative output of the disk-stream interaction.

The flare in the well-known AGN 1ES\,1927+654 \citep{Trakhtenbrot2019_1ES1927} exhibited an optical light-curve reminiscent of a TDE in rise time, peak luminosity, and decline rate. However, there were no telltale TDE features in the optical/UV spectroscopy, which instead revealed a changing-look AGN event occurring on a timescale of months. 
A follow-up study \citep{Ricci2020_1ES} speculated that the disappearance and reappearance of the X-ray emitting corona (also on timescales of months) was indeed linked to the tidal disruption of a star onto the pre-existing AGN accretion disk. 
This interpretation was based on qualitative agreement with models \citep{Chan19}, which were limited in their predictive power \cite[see more recent progress in][]{Chan2020}.

Another complex nuclear transient that occurred in a known AGN is CSS100217:102913+404220, where the SDSS spectrum prior to the flare shows signatures of a NLSy1 galaxy \citep{Drake:2011a}. 
\citet{Drake:2011a} rule out a TDE based on several factors: the light curve evolution does not follow a $t^{-5/3}$ decay, the peak brightness ($M_{V, CSS} = -23$) is much higher than usually observed in TDEs ($M_{V} \sim -20$; see \optchap), and the fitted temperature is too low ($T = 1.5 \times 10^{4}$ K) compared with theoretical expectation ($T\simeq10^{5}$ K). 
The light curve and the evolution of the narrow Balmer lines in the optical spectrum are consistent with a Type IIn supernova.

As in the case of AT 2017bgt-like events, the nature of this transient is hard to interpret. While CSS100217 could be a nuclear Type IIn SN or AGN flare, its fitted temperature is consistent with values seen in optical/UV TDEs (see \optchap), where the emission may be formed in outer shocks or reprocessing material rather than in a directly visible accretion disk. 
\citet{Drake:2011a} point out that other NLSy1 galaxies do not show the same level of optical variability and that the increase in the narrow H$\alpha$ line strength after the flare died away occurred on a timescale too short to originate in the narrow line region. 
A focus on following up nuclear transients in AGN hosts, particularly in NLSy1 galaxies, and on quantifying the rates and range of TDEs and nuclear SNe in a control sample of quiescent galaxies, will reveal the connection between AGN and TDEs, whether it be physical or just due to mis-classification of AGN flares and/or nuclear SNe as TDEs.

\subsection{Summary}\label{sec:AGN_summary}

Unambiguous TDE classification remains challenging, as any single observed property may be consistent with that of a persistent or flaring AGN. 
However, a constellation of unusual features like those cited in the previous sections and their consistency with rough expectations from theory argue that at least some TDE candidates, even unusual ones like  ASASSN-15lh \citep{Leloudas:2016a} and PS1-11af \citep{Chornock:2014a},
may in fact be tidally disrupted stars.

The observed overlap of TDE and AGN properties, particularly in the case of highly-variable AGN and AGN disk instability induced flares, complicates TDE classification
and highlights the importance of archival data in constraining the level of SMBH activity prior to and long after a transient's detection. Known AGN should not be excluded from transient searches. It would be helpful
to make use of existing datasets and related AGN selection criteria including:
{\it WISE} for MIR-based AGN selection \cite[e.g., following the criteria in][]{Stern2012_MIR_AGN_WISE} and variability;
{\it ROSAT} (and even {\it Swift}/BAT) for X-ray luminosity and X-ray SED shape; FIRST/NVSS for radio-based discrimination between galaxy star formation and AGN activity \cite[e.g., following the SFR prescriptions of][]{Hopkins2003_SFR}; SDSS, 2dF/6dF, and earlier spectroscopic surveys, as well as relatively new efforts like GAMA and OzDES. All can be used to determine whether the transient in question is an unusual AGN or perhaps a TDE.

AGN and TDE may also be linked physically.
New discoveries of dramatic, UV/optical-bright flares from accreting SMBHs, which share some TDE characteristics, raise
the possibility that tidal disruption phenomena may
occur in existing AGN and lead to
extreme AGN variability. From a theoretical point of view, it is not yet clear how often to expect a tidal disruption event in a pre-existing AGN.
Nor do we understand why some TDEs and flaring AGN both have Bowen fluorescence lines.
Observationally, we do not know at present whether TDEs in pre-existing AGN are more common than those associated with dormant SMBHs or how to cleanly distinguish TDEs from the  flares of known AGN.

\section{Distinguishing TDEs from SNe}
\label{sec:TDEvSNe}

At first sight, optical/UV TDEs (e.g., PS1-10jh; see \optchap) can have observed properties similar to those of core collapse (CC) SNe: light curves with peak luminosities between those of ``normal" and superluminous (SL) SNe \citep[e.g.,][]{Arcavi:2014a}, light curve timescales like those of SLSNe \citep[e.g.,][]{gal-yam2012}, and broad emission line features.
Indeed, ASASSN-15lh remains a debated TDE versus SN case (see Section \ref{sec:oneoffs}). However, for most TDEs, a closer look at their properties reveals stark differences with SNe.

\subsection{Summary of Observational Distinctions}

The events in the PS1-10jh-like class of optical/UV TDEs (see \optchap) have the following properties that are not typically seen in SNe:

\begin{enumerate}
\item Blue, constant colour for hundreds of days;
\item Very late-time (years) UV detections;
\item Broad emission line profiles with no blueshifted absorption;
\item Strong, broad He~II;
\item No late-time narrowing of emission lines in nebular phase;
\item Accompanying X-ray flare (in some cases).
\end{enumerate}

\subsection{Temperature Evolution}

\begin{figure}
	\begin{center}
		\includegraphics[width=0.7\textwidth]{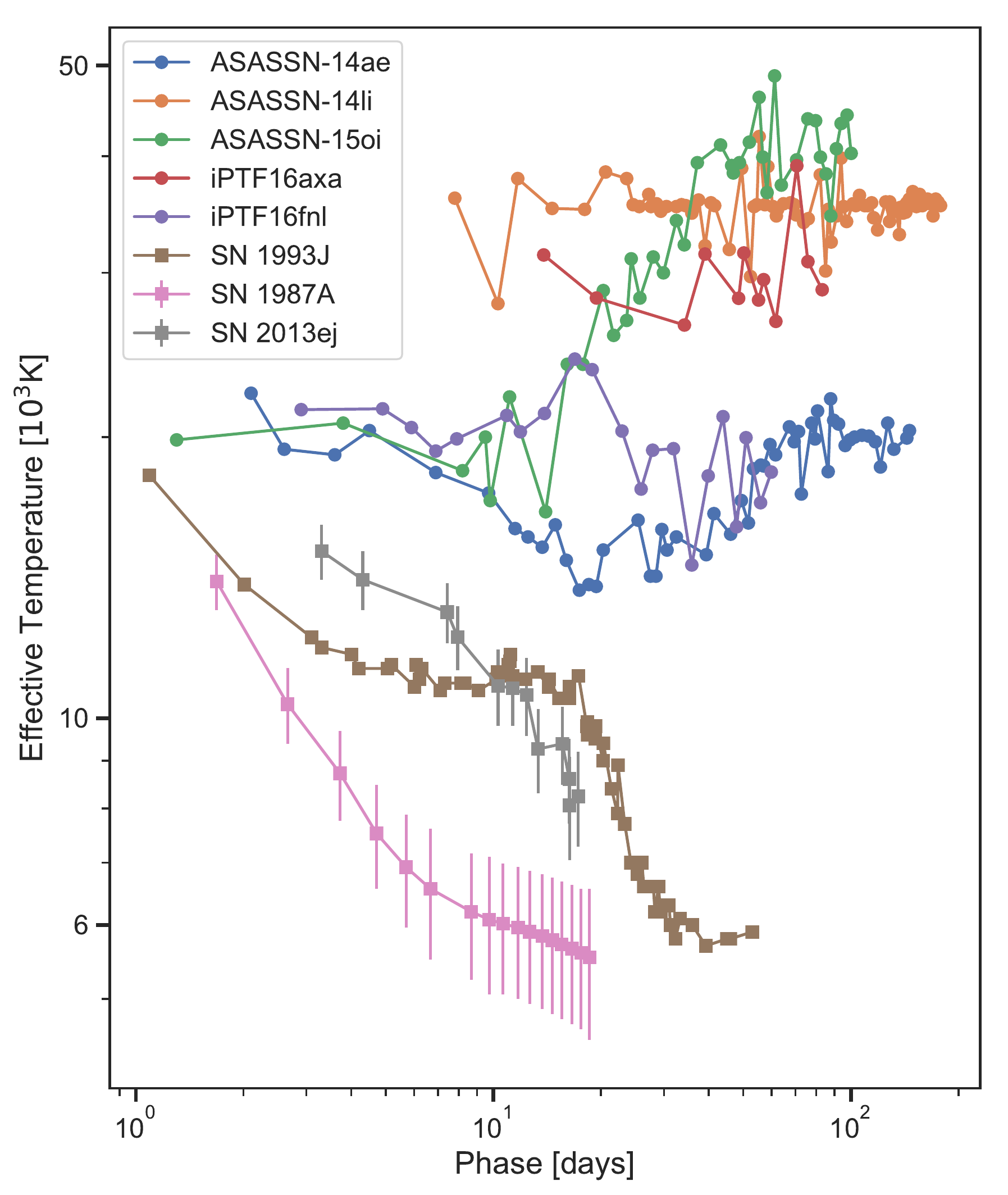}
		\caption{Effective temperature evolution for optical/UV TDEs \cite[circles, in days from discovery;][]{Hung2017,Holoien:2014a,Holoien:2016a,Holoien:2016b} and for hydrogen-rich core collapse SNe \citep[squares, in days from explosion;][]{Menzies1987,Richmond1994,Valenti2014}. TDEs remain hot, while SNe cool within a few weeks. H-stripped SNe, which are not shown, cool even faster.\label{fig:tde_sn_tempr}}
	\end{center}
\end{figure}

The colour of optical TDEs, implying an effective temperature of a few $\times10^4$K, remains constant for weeks or even months (see \optchap\ and references therein). While such temperatures are observed in SNe at early times, they cool within days to much lower temperatures (see Fig. \ref{fig:tde_sn_tempr}). 

In addition, some TDEs show continued UV emission years after discovery \citep{vVelzen2019}; such long-lasting UV emission is not seen in SNe.

\subsection{Spectral Line Profiles}

Most SNe spectra display lines with P-Cygni profiles, which originate in expanding ejecta. The line profiles of optical/UV TDEs are very different, showing no absorption and sometimes asymmetric emission profiles (see \optchap\ and Figure \ref{fig:tde_sn_spec}). Even Type IIL SNe, which show weaker P-Cygni absorption \citep[e.g.,][]{Gutierrez14} compared to Type IIP SNe, are still not as emission-dominated as TDEs. 

\subsection{Spectral Line Species}

\begin{figure}
	\begin{center}
		\includegraphics[width=0.65\textwidth]{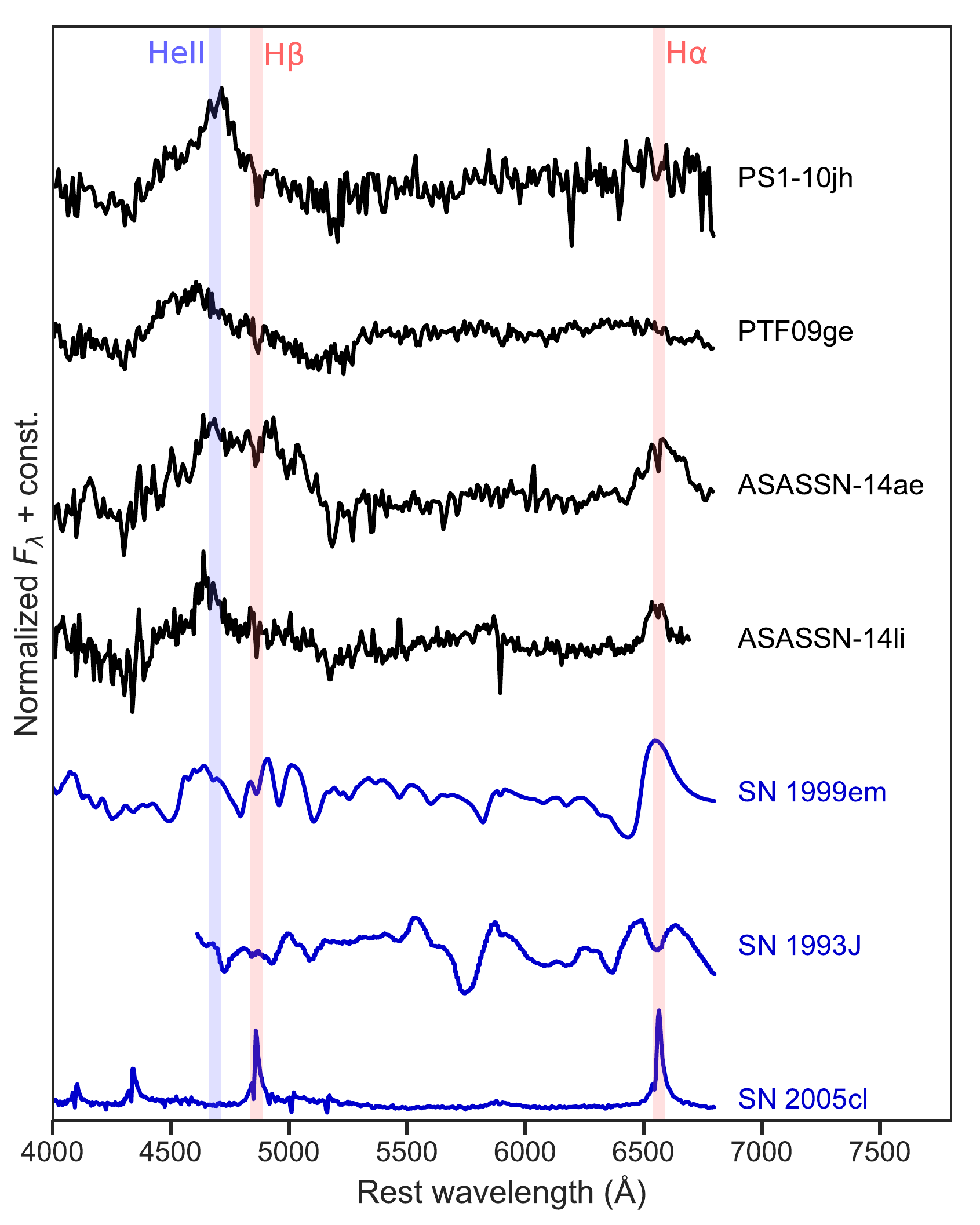}
		\caption{Continuum subtracted spectra of optical/UV TDEs \citep[black;][]{Gezari:2012a,Arcavi:2014a,Holoien:2014a,Holoien:2016b} and of different subtypes of hydrogen-rich core collapse SNe \citep[blue; from bottom: IIn, IIb and IIP;][]{Kiewe2012,Barbon1995,Leonard2002}. Both the spectral species and line profiles are different between TDEs and SNe.}
		\label{fig:tde_sn_spec}
	\end{center}
\end{figure}

The spectral line species in optical TDEs differ from those of any known SN. Specifically, broad \heii\ is not seen in any SN type at a strength comparable to H, in contrast to the broad \heii\ and Balmer lines of some optical TDEs (Fig. \ref{fig:tde_sn_spec}). Ca features that are ubiquitous in SNe are not seen in TDEs at all, while only a few TDEs have shown Fe features, another element that is commonly observed in SN spectra. 

\subsection{Spectral Line Width Evolution}

The spectral lines of optical TDEs remain relatively broad even at late times (months to years after peak); in contrast, SN spectra become ``nebular" on those time scales, revealing only narrow emission lines from inner slow moving material as the ejecta expand and become transparent.

\subsection{X-Ray Emission}

Some optical TDEs are accompanied by X-ray emission, as expected from newly formed accretion disks. Most SNe, on the other hand, do not 
emit in X-rays, except for cases of obvious interaction between the SN ejecta and dense circumstellar material (CSM). Furthermore, in those cases, narrow emission lines from the unshocked CSM dominate the spectra, in stark contrast to the broad lines seen in the main class of optical TDEs.


\subsection{Summary}

Several observational properties, which are readily measured for transients, help distinguish TDEs from SNe. This, in addition to the ample knowledge of SN populations and their typical emission properties, is the reason that there is relatively little confusion between these types of events. The notable exception is the case of ASASSN-15lh, which is a clear outlier to both known TDE and SN populations. Analysing the location of events with respect to the host galaxy centre may provide clarity; if future 15lh-like events are all found in the host centres, the TDE interpretation will be strengthened, else these sources are likely SNe.

\section{Distinguishing Jetted TDEs from GRBs}


The discovery of $\gamma$-ray bursts (GRBs) of cosmic origin were reported by \citet{Kleb:1973}. GRBs were a bi-modal distribution and could be divided into two categories based on the duration of their prompt emission: short and long \citep[e.g.,][]{Kouveliotou:1993}. Short GRBs (SGRBs) have prompt emission (T$_{90}$) lasting approximately 2 sec or less, while long GRBs (LGRBs) have prompt emission from about 2 up to several thousand seconds.  A more recently discovered class of ``ultra-long" GRBs (GRB~111209A, GRB~121027A, and GRB~101225A) have prompt emissions at high energies from a few hundred to a few thousand seconds \citep[e.g.,][]{Levan:2014}. 

Jetted TDEs became a new category of TDEs after the unusual discovery of GRB~110328A by \emph{Swift}'s $\gamma$-ray telescope \citep[hereafter, Sw~J1644+57; ][]{Burrows:2011a,Zauderer:2011a,Bloom:2011a,Levan:2016}. The same year, \emph{Swift} discovered another event, Sw~J2058+05
\citep{Cenko:2012b}, with very similar properties.  In this section, we briefly describe the discovery of jetted TDEs and the primary ways in which they can be differentiated from GRBs via 1) prompt emission energy levels, 2) timescales and light curve; and 3) host galaxy association and the location of the transient within.  More details on hard X-ray and $\gamma$-ray selected TDEs are described in the \gammachap.

\subsection{Prompt Emission, Timescales, and Light Curve}

\begin{figure}
	\begin{center}
		\includegraphics[width=1.0\textwidth]{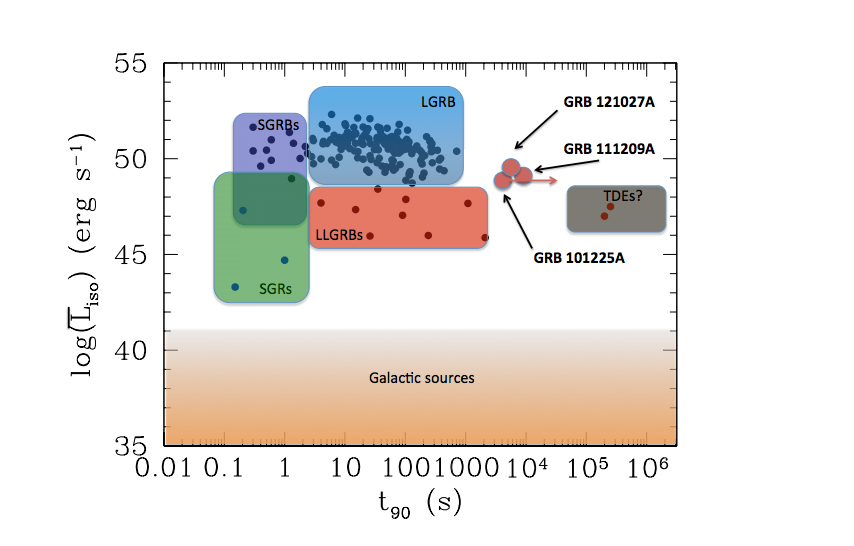}
		\caption{Duration of burst versus approximate average luminosity over that duration for transients in the $\gamma$-ray sky. The figure compares the properties of soft-gamma repeaters (SGRs) in our own Galaxy, long- and short-duration GRBs (LGRBs and SGRBs), low-luminosity GRBs (LLGRBs), three GRB outliers (GRB 101225A, GRB 111209A, and GRB 121027A), and two very long transients thought to be jetted TDEs. Compared to GRBs, jetted TDE candidates exhibit high amplitude flaring at early times and prompt emission in X-rays lasting for weeks, i.e., $t_{90}$ of $10^5$-$10^6$~s.
		Figure reproduced from \citet{Levan:2014}.}
		\label{fig:TDEvGRB}
	\end{center}
\end{figure}

Jetted TDEs differ from GRBs in several ways, including high amplitude flaring at early times and prompt emission in X-rays lasting for weeks, i.e., $t_{90}$ of $10^5$-$10^6$~s (Figure \ref{fig:TDEvGRB}). 
The X-ray luminosity is also much greater \citep[see, for example, Figure 4 of][]{Cenko:2012b}.  Additionally, for jetted TDEs, the X-ray light curve (after initial flaring) falls off as roughly $t^{-5/3}$.  Sw~J1644+57 and Sw~J2058+05 have similar properties, including a X-ray light curve that slowly declined as $t^{-5/3}$ and then abruptly cut off, suggesting that the relativistic outflow shut off; see Figure 4 of \citet{Zauderer:2013} and Figure 1 of \citet{Pasham:2015a}.

\subsection{Relationship to Host Galaxy}
Another approach to distinguish jetted TDEs from GRBs is to evaluate the environment, and, if detectable, the type of host galaxy and where the transient lies within it. Long GRBs were determined to come from the deaths of massive stars based on their association with star forming galaxies, with the star-forming regions in those host galaxies, and, more specifically, with SNe themselves \citep[see][and references therein]{Hjorth:2012}.
Studies of the host environment \citep[see][and references therein]{Fong:2015} also contributed to the model of short GRBs as originating from compact object mergers, for which the discovery of GW170817 \citep{PhysRevLett.119.161101} has now provided the strongest evidence. 
On the other hand, the jetted TDE candidates observed to-date have radio emission consistent with arising from the centres of inactive galaxies.
For example, VLBA observations localised the radio afterglow of Sw~1644+57 to the host galaxy centre \citep{Zauderer:2011a}; multi-wavelength observations were important to 1) localise the afterglow, as the error circle for the $\it{Swift}$/XRT was more than an arcsecond; 2) obtain redshift information (e.g., from optical spectroscopy); and 3) classify the host galaxy (especially given that there were no previous observations). 

\subsection{Summary}
For current and future wide-field surveys (e.g., VLA Sky Survey; VLASS), \citet{Metzger+:2015} discuss how various types of transients may be distinguishable based on emission timescales and energies.  A combination of prompt emission timescales, light curves, event energetics, host galaxy properties, and transient location relative to the host centre will help to distinguish future TDEs from GRBs.  In a case where the transient is detected after the high-energy emission has faded, the work may be more difficult. However, host galaxy properties will still be useful for population-wide, statistical studies.


\section{Other Potential TDE Impostors}

\subsection{Stellar Collisions in Galactic Nuclei}

Galactic nuclei hosting SMBHs are some of the most dense and collisional environments in galaxies, with stellar densities of $10^6$-$10^7$ pc$^{-3}$ \cite[see][for a review]{Neu+20}. Stellar collisions and mergers could thus be abundant there \cite[see][and references therein]{Dal+09}, making explosive transient events far more likely. Because the densest nuclear star clusters are associated with galaxies hosting relatively low mass ($< 10^8$ M$_\odot$) SMBHs, 
stellar collisions may generate transient events in nuclei of similar masses to those expected to produce TDEs, leading to misidentifications.
Moreover, mergers of binary stars can be induced by secular evolution in triple systems \citep{Perets09c}; in galactic nuclei, every binary forms a triple with the SMBH and is therefore potentially sensitive to such secular evolution. In other words, binary mergers could be triggered near SMBHs \citep{Antonini10b,Antonini:2012,Prodan:2015a,Stephan19}, potentially mimicking TDEs.

Nevertheless, there are likely differences between TDEs and nuclear stellar collisions and mergers that could distinguish their populations individually and/or statistically. For example, mergers/collisions may produce reddened transients, perhaps similar to the merger in the V1309 Scorpii system \citep{Tylenda2011}. More generally, 
stellar collisions/mergers should be abundant in other dense stellar environments, such as globular clusters, so their their expected spatial distribution and properties would need to be reconciled with the nuclear (by definition) environments of TDE candidates.

Another possibility is that stellar collisions and mergers near a SMBH 
could remove material from stars that is then accreted by the SMBH. For example, the rate of grazing collisions can be enhanced through the capture of stars inspiraling to the SMBH due to gravitational wave emission \citep{Metzger:2017}. Such grazing collisions can strip material from the stars, which would later be accreted by the SMBH or accumulate to form a disk around the SMBH and eventually produce a flare. 

Other types of induced mergers and collisions near SMBHs could produce gas clouds that later accrete onto the SMBH and generate flares (e.g., SMBH-induced binary mergers, \citealt{Antonini:2012, Prodan:2015a}). Such events might impersonate TDEs or disk instability flaring. It is not clear, however, why such induced stellar mergers should be preferentially observed in the hosts most favoured by TDEs, i.e., quiescent, Balmer-strong (QBS) galaxies and the subset of post-starburst (PSB, or ``E+A") galaxies (\citealt{Arcavi:2014a, French:2016a, graur2018}; see the \hostchap).

\subsection{Micro-TDEs}
The tidal disruption of stars by stellar-mass black holes potentially gives rise to a different class of transients. Such ``micro-TDEs," especially when occurring in or near galactic nuclei, could be incorrectly interpreted as TDEs associated with the central SMBH. 
\citet{Perets:2016a} suggested that micro-TDEs arising from the disruption of MS stars or planets may appear as ultra-long GRBs ($\sim 10^3$ to a few $\times$ $10^4$ s). The timescales could be longer for evolved star progenitors, whose disruption might produce longer-term transient X-ray sources. Such micro-TDEs could masquerade as jetted TDEs detected in X-rays and $\gamma$-rays.

While micro-TDEs should occur across different environments, including dense non-nuclear stellar clusters and even in the field \citep{Perets:2016a}, 
they could also be frequent in galactic nuclei, given the highly collisional environment there (as discussed in the previous section). 

\citet{lin2018} reported a transient event in a star cluster offset by 12.5 kpc from the centre of its host galaxy ($z=0.055$). Observational signatures of this event are similar to those of TDE candidates: the light curve decays by $t^{-5/3}$ over 10 years, the X-ray spectrum is soft and well described by a thermal accretion disk that cools with decreasing luminosity, and the peak luminosity of the flare is 10$^{43}$ erg s$^{-1}$. This event may be a micro-TDE of a post-MS star.  Another possibility is that the TDE arose from a star shredded by an intermediate-mass black hole in the star cluster, which would represent a new probe of that elusive black hole population.

\subsection{Circumbinary Accretion Flows}
Many studies \cite[see][for a review]{DeR+19} explore the possibility of binary massive black holes accreting from a circumbinary disk \citep[e.g.,][and references therein]{Iva+99,Mac+08,Cua+09,Shi+12,Dor+13,Tanaka:2013a,Mun+20, Rag+20}.
\citet{Tanaka:2013a} suggests that a SMBH binary (SMBHB) can clear a central cavity in its accretion disk, which then appears dimmer and softer than an equivalent single AGN disk. Gas streams of 0.1 solar masses, which are intermittent on timescales depending on the binary period, then infall towards the SMBHB from the cavity’s lip. The streams then shock, giving rise to a TDE-like flare. If the recurrence period is long compared to the era of
transient monitoring, a single detection might be misidentified
as a TDE. 

The recurrence period, debris velocities, and energy yields
in a cavity flare depend on the SMBHB properties and need not correspond to the predicted tidal radii of stellar disruptions. Thus, the timescales for such flares might not be directly related to SMBH mass (cf. \citealt{vVelzen2019}). Subtler details might distinguish binary cavity flares: the steepness of their fading or their occurrence around a SMBH that is too massive for a TDE. For example, \citet{Leloudas:2016a} and \citet{Kreuhler2018} infer $M_{BH}>10^8$ M$_{\odot}$ for the weak-lined TDE candidate ASASSN-15lh.
If SMBH binaries are a common product of
the galaxy-galaxy mergers that produce post-starburst galaxies, then the cavity flare mechanism, like stellar tidal disruption, may favour such host galaxies; the expected rate is currently difficult to assess.

\section{Distinguishing TDE Candidates Statistically}

The demographics of TDE-candidates, e.g., their frequency, spatial offset from the host galaxy nucleus, and host properties, may provide evidence for their existence by {\it statistically} distinguishing them from impostors.
The expected rates of TDEs are reviewed in the
\ratechap, and TDE host galaxy properties in the \hostchap.
Here we discuss statistical arguments suggesting that at least a fraction of TDE candidates are real, while some individual TDE detections remain ambiguous. Specifically, we compare the TDE nuclear offsets and the SMBH masses, stellar masses, and star formation histories of TDE hosts to those of AGN and SNe.

\subsection{Nuclear Offsets}

\begin{figure}[!ht]
	\begin{center}
		\includegraphics[width=0.75\columnwidth]{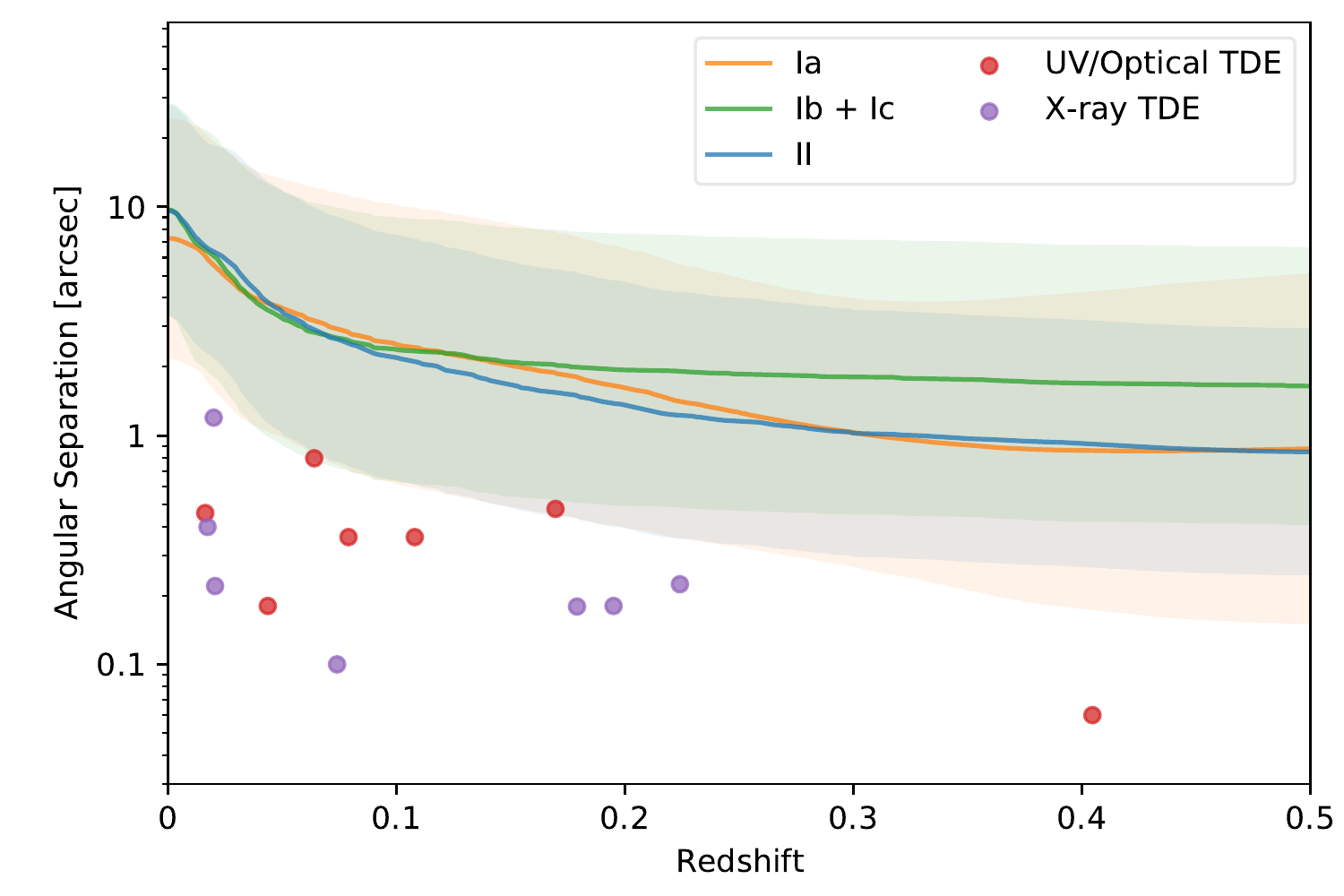}
		\caption{Angular separation from host galaxy centre versus redshift for UV/optical and X-ray TDE samples (red and purple points, respectively) and for SNe (coloured mean lines and $\pm 1\sigma$ bands). The data are drawn from the transient-host pair catalogue of Qin et al. (in prep.). For TDEs, the angular separation uncertainties are generally unrecorded in the literature and probably dominated by the uncertainty in the position of the host's centre, especially at lower redshifts.
		At all redshifts, the SNe tend to lie further from the centre than TDEs, with angular offsets of over 1 dex ($1.8\sigma$) larger. This difference may arise in part from a selection effect, as certain types of SNe are likely to occur near the dense centre, but are harder to detect there, and transients may be classified as TDEs after consideration of their proximity to the nucleus.}
		\label{fig:offsets}
	\end{center}
\end{figure}

SNe in general have higher observed rates than TDEs (see \ratechap), and most observed SNe are significantly offset from the galactic centre (Figure \ref{fig:offsets}). However, these offsets are in part a selection effect; intrinsically, some types of SNe are more likely than others to occur near the centre. While TDEs are nuclear sources by definition (except for presumably rare cases around a recoiling BH), SNe generally follow the mass distribution of their progenitor stellar population in the host galaxy. Because TDE candidates significantly differ from typical SNe, it is hard to explain all TDE candidates as SNe, as those SNe masquerading as TDEs would need both to occur in or very near the nucleus and be unusual. Although the environments of galactic nuclei are unique, it is challenging to envision them giving rise to SNe that are rare elsewhere. 

Improving the statistics of centrally located SNe requires uniform, higher spatial resolution surveys. Arguably the highest spatial resolution transient survey currently running is \emph{Gaia}, with a typical astrometric accuracy of a few tenths of an arcsecond. \citet{blagorodnova2016} show that a transient fainter than $G = 16$ mag (roughly equivalent to $V = 16$) can be resolved by \emph{Gaia} if it is more than $0.2$ arcsec away from the bulge of the galaxy. Within this $0.2$ arcsec volume, they find that SNe and TDEs occur in roughly equal numbers. Further improvements in \emph{Gaia} astrometry may push this ratio in favour of TDEs (see also \citealt{Kostrzewa18} for a discussion on nuclear transients in \emph{Gaia}).

Connections between extreme SNe and TDE claims \cite[e.g.,][]{Komossa2009_J0952,Drake:2011a} are questioned by \cite{Gezari:2009a}, see Section \ref{sec:TDEvSNe} above.
Nevertheless, unusual conditions in the galactic core, including a dense ISM in the circumnuclear region of a starved/dormant AGN, may boost peculiar SNe rates there. Gas confinement may alter SN shock physics and radiative efficiency, potentially producing an abnormally shallow light curve \citep{Saxton:2018}. 
A larger and more representative census of nuclear SNe, as well as improved models, is needed to resolve this issue.

\subsection{Host Galaxy SMBH and Stellar Masses}
\label{host_galaxy_masses}

\begin{figure}[!ht]
	\begin{center}
		\includegraphics[width=0.75\columnwidth]{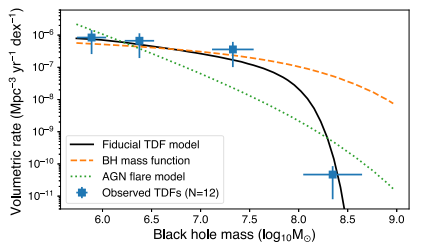}
		\caption{Differences in expected black hole mass function for TDEs vs. AGN from \citet{van-Velzen:2018a}. The green dotted line represents the expected volumetric rate of AGN flares, assuming that the flare reoccurrence time scales linearly with M$_{\rm BH}$. This model fails to reproduce the observed volumetric rate of observed TDE candidates. Instead, a simple model where the SMBH mass function is multiplied by a constant TDE rate (orange dashed line), and that takes into account the suppression due to direct captures, provides a better description of the observations (black solid line).}
		\label{fig:vv2018}
	\end{center}
\end{figure}

TDEs prefer hosts with 
SMBH masses $M_{\rm BH} < 10^8\,M_{\odot}$ \citep{Wevers:2017a, 2019MNRAS.487.4136W, Mockler:2019}. 
This is not surprising, given that the tidal radius should be outside the Schwarzschild radius for TDEs to be observable.
TDEs and AGN can be distinguished statistically using differences in their black hole mass functions \citep{van-Velzen:2018a}.
The expected volumetric rate of AGN flares, assuming that the flare reoccurrence time scales linearly with $M_{\rm BH}$, does not match the observed volumetric rate of observed TDE candidates (Figure \ref{fig:vv2018}). Instead, a simple model where the SMBH mass function is multiplied by a constant TDE rate, and that takes into account the suppression due to direct captures, is more successful. \citet{van-Velzen:2018a} argue that the strong suppression of the volumetric rate for $M_{\rm BH} > 10^8\,M_{\odot}$ can only be explained by the TDE scenario.

At least some AGN that arise from lower mass SMBHs may be distinguished from the TDE population by other signatures. 
For example, AGN driven by SMBHs with
$10^6 \lesssim (M_{\rm BH}/M_{\odot}) \lesssim$ few $\times 10^7$ often produce relatively narrow Balmer lines (${\rm FWHM}\sim1200$-$2000\,{\rm km\,s}^{-1}$) and thus are classified as Narrow Line Seyfert 1 galaxies (NLSy1s). 
Such optical spectra differ from what is seen in most (optical) TDE candidates, whose lines are much broader ($\gtrsim15000\,{\rm km\,s}^{-1}$).  Because the
\HeIIop\ line width is similar to those of the Balmer lines in AGN, it would be considerably more narrow than in TDEs.

\begin{figure}[!ht]
	\begin{center}
		\includegraphics[width=0.75\columnwidth]{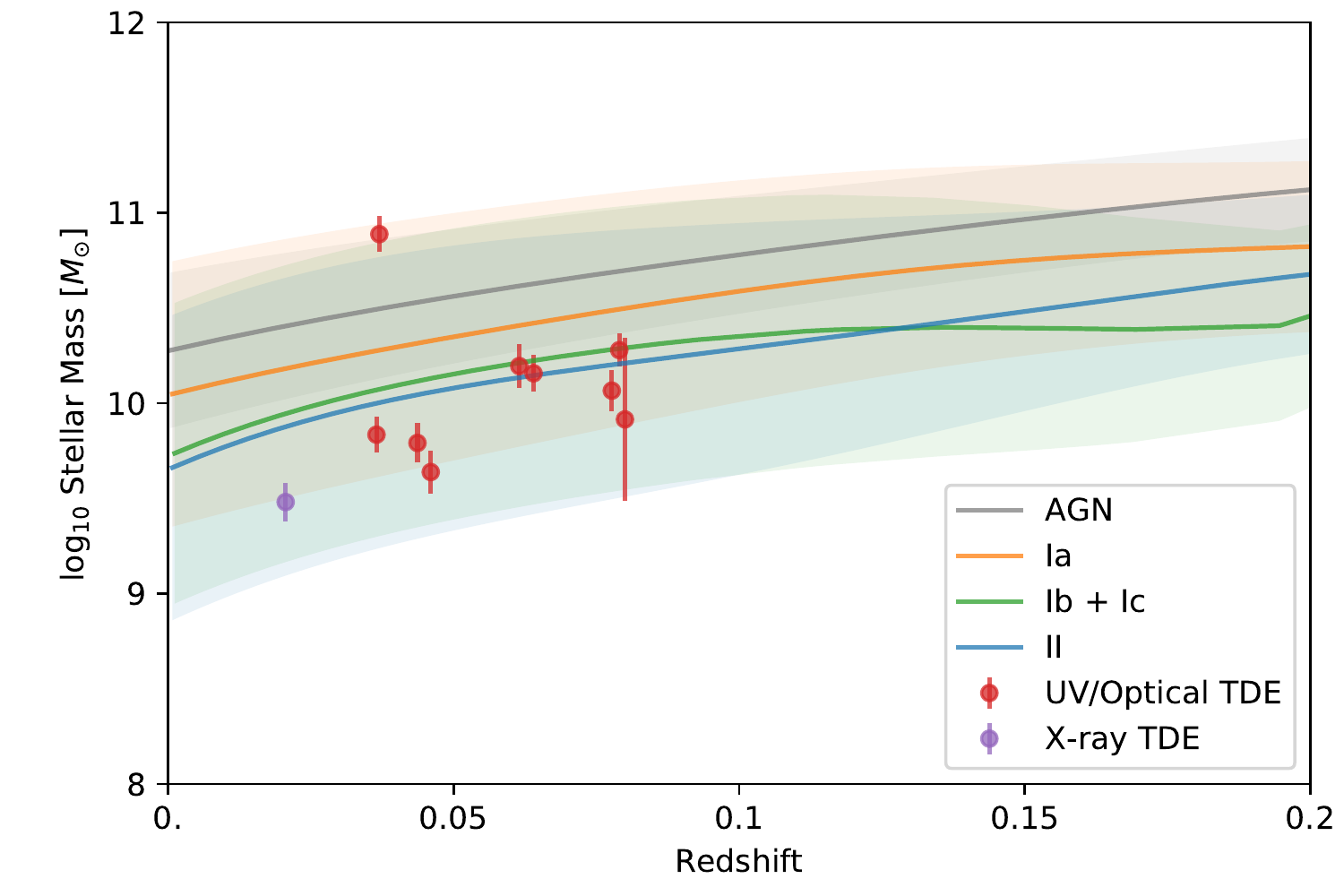}
		\caption{Host galaxy stellar masses versus redshift for UV/optical (red points)
		and X-ray (purple point)
		TDEs, SNe (orange, green, and blue mean lines and $\pm 1\sigma$ bands), and AGN (grey mean line and $\pm 1\sigma$ band). Data are compiled from the transient-host catalogue of Qin et al. (in prep.). 
		The high outlier event, AT2018dyk (ZTF18aajupnt), is listed as an optical/UV TDE in the \hostchap. 
		For its stellar velocity dispersion of 
		$121 \pm 3.8$ km s$^{-1}$ (SDSS DR8 spectral catalogue; \citealt{2011ApJS..193...29A}), its SMBH mass 
		would be roughly $1.6\times10^7$ M$_{\odot}$ (from the scaling relation in \citealt{2011Natur.480..215M}), only marginally higher than typical of other TDE host galaxies (\citealt{2019MNRAS.487.4136W}). 
		\citet{2019ApJ...883...31F}
		alternately classify this source as a rapidly brightening LINER, 
		whose later UV and optical spectra reveal a NLS1 with strong coronal lines.
		%
		The stellar masses of these TDE host galaxies are, on average, 0.74 dex ($\sim$1.8$\sigma$) below those of AGN hosts at the same redshift.
		}
		\label{fig:qin2019}
	\end{center}
\end{figure}

By extension, the host galaxies of TDEs and AGN might be expected to have different stellar mass distributions. Data for TDE, AGN, and SNe hosts suggest that TDE host stellar masses tend to be smaller than those of AGN (Figure \ref{fig:qin2019}).
This offset is also apparent if we consider only the TDEs that satisfy most of the criteria laid out in this chapter and whose hosts have known stellar masses, i.e., the 14 TDEs with broad optical H and He emission lines and the two objects with the most robust classification of ``X-ray" TDEs by \citet{2017ApJ...838..149A}, as in the \hostchap.

Furthermore, the host properties of X-ray TDE candidates grouped by the strength of the TDE claim (\hostchap) indicate that the least certain TDEs (those classified as ``possible TDEs" by \citealt{2017ApJ...838..149A}), have on average higher SMBH masses and brighter host galaxy absolute magnitudes than ``likely" or ``X-ray" TDEs. While robust conclusions are limited by the small sample sizes, this result suggests that the ``possible" TDEs are a different population, e.g., misidentified AGN flares,
which tend to inhabit more massive and brighter host galaxies.
For such flares, the lack of high quality data leads not only to the ``possible" TDE label, but also makes them appear similar to TDEs.

\subsection{Other Host Galaxy Properties}
\label{host_galaxy_prop}

In addition to SMBH mass (and, by correlation, total stellar mass), other host properties can be used to distinguish TDEs statistically.  TDEs prefer hosts with unusual, A-star dominated spectra, i.e., quiescent, Balmer-strong (QBS) galaxies and the subpopulation of post-starburst (PSB or ``E+A") galaxies (\citealt{Arcavi:2014a, French:2016a, graur2018,LawSmith2017}; 
see also the \hostchap).
The quiescent galaxies common among TDE hosts generally lack strong H$\alpha$ emission, although several show broad emission lines indicative of Type 1 AGN. A higher fraction of TDE hosts exhibit
weak, narrow emission lines, with line ratios indicating ionisation from sources other than star formation. Could the observed TDE population arise from the tail end of Type 2 AGN variability? We address this question here statistically.

The recent star formation histories of TDE host galaxies are different on average than Type 2 Seyfert or LINER AGN. 
For galaxies in the SDSS main spectroscopic sample that 1) have stellar masses typical of TDE host galaxies ($\log\left(M_\star/M_\odot\right)\simeq 9.5-10.5$) and 2) have all four emission lines required for AGN classification on the BPT diagram \citep{bpt}, 92\% are star forming, 5\% are Seyferts, and 3\% are LINERs, according to the \citet{kauffmann} criteria. While between 36-75\% of TDE candidates with broad H/He lines are found in QBS galaxies, only 9\% of Seyferts and 17\% of LINERs are. If TDEs were the tail end of the Type 2 AGN LINER or Seyfert distributions, we would expect that 9-17\% of the host galaxies were QBS galaxies, instead of 36-75\%. Thus, it is unlikely that most of the observed broad H/He line TDEs are caused by the tail end of normal Type 2 AGN variability. 

Furthermore, many galaxies with LINER-like spectra maybe be ionised by sources other than low-luminosity AGN, weakening the possibility that the observed TDE population is caused by extreme Type 2 AGN variability. Many LINER-like galaxies identified using the BPT diagram could be ionised by merger-induced shocks \citep{Rich2015} or post-AGB stars \citep{Yan2012}. Both of these possible ionisation sources are expected to occur during the QBS or PSB stage, so the proportion of QBS and PSB galaxies hosting a true low-luminosity AGN will be lower than that inferred from the BPT diagram alone.

The host galaxies of the main optical and X-ray TDE classes are also distinct from those of core-collapse (CC) SNe. CC SNe originate from massive stars and thus are found almost exclusively in star forming regions. Star formation mostly ended more than $\sim$100 Myr ago in QBS galaxies, the favoured hosts of TDEs, so massive stars, and thus CC SNe, in QBS galaxies are unlikely.

\section{Conclusions and the Future} \label{sec:conclusions}

While TDE classification remains ambiguous---any single observed property may be found in other types of transients---a constellation of unusual features like those cited above, consistency with rough expectations from TDE theory, and the statistics of the transient and host galaxy populations argue that at least some candidates do in fact arise from stars tidally disrupted by SMBHs.  
The tallies of potentially distinguishing features included in this chapter are not meant to be rules, but initial guidelines, for these early days of TDE detection. 
In the RubinObs/LSST, SDSS-V, and \emph{eROSITA} era, new, more extreme types of AGN and other transients will be discovered that are likely to cast doubt on some current TDE claims and to further complicate future TDE classifications. 
Yet, over the same period, thousands of new TDE candidates, as well as improvements to theoretical models of stellar disruption and disk accretion by SMBHs, will help us define any unique combinations of TDE signatures.

Some of the key outstanding questions and future needs are:
\begin{itemize}
    
    
    \item Are TDEs are significantly less absorbed compared to AGN or is this difference a selection bias?
    
    \item Why do the X-ray spectra of some thermal-dominated TDEs harden as they evolve, while others do not?
    
    \item What is the physical driver of the hyper-variable and ``changing-look" AGN now detected in time-domain surveys? This question is relevant here because flares from specific changing-look AGN (e.g., SDSS J015957.64+003310.5; \citealt{lamassa2015}) have been attributed to the tidal disruption of a star around a previously active SMBH \citep{Merloni:2015a}. On the other hand, changing-look AGN may just be extreme examples of regular, continuous AGN variability.  Thus, understanding the demographics and drivers of both changing-look and hyper-variable AGN may help to determine characteristics that distinguish them from TDEs. A crucial first step would be a large, fair census of such objects.
    
    \item What other kinds of flares in persistent AGN are out there and how are they related to TDEs? 
    
    
    \item What signatures can distinguish between AGN flares due to variability in a pre-existing accretion disk and those due to TDEs?
   
    \item Are TDEs within pre-existing AGN more common than those around dormant SMBHs, suggesting a link between AGN and TDEs or similarities in the conditions that produce them? 
    
    
    \item What does the detection of Bowen fluorescence lines, indicative of extreme UV radiation fields, in many TDE candidates and in flaring AGN tell us about the physical links between AGN and TDEs?
    
    \item To strengthen the TDE interpretation, it is critical to find additional examples of one-off events like ASASSN-15lh. If similar sources all lie in the host galaxy centres, then the case for TDEs is stronger. Otherwise, one example detected off-nucleus would suggest a new kind of superlumious SN.
    
    \item Are there other potential TDE impostors, such as stellar collisions, ``micro-TDEs," or circumbinary flares, that occur more frequently than TDEs in quiescent, Balmer-strong galaxies? Or in the ``post-starburst" subclass of these galaxies?
    
    \item Do any known SNe types or other potential TDE impostors occur more frequently than TDEs in galactic nuclei?
    
    \item What is the relationship between TDE claims and the high proportion of LINERs also found in post-starburst galaxies?

    
\end{itemize}

\begin{acknowledgements}
The authors thank ISSI for their support and hospitality and the review organisers for their leadership in coordinating these reviews. This work was performed in part at the Aspen Center for Physics, which is supported by National Science Foundation grant PHY-1607611, during the January 2019 Aspen conference on \emph{Using Tidal Disruption Events to Study Super-Massive Black Holes}. We are grateful to Sjoert van Velzen and Ryan Foley for leading a discussion there about possible differences between TDEs and supernovae. We also thank Nicholas Stone, Sixiang Wen, and Dennis Zaritsky for helpful information.
AIZ acknowledges support from grant HST-GO-14717.001-A
from the Space Telescope Science Institute (STScI), which is operated by the Association of Universities for Research in Astronomy (AURA), Incorporated, under NASA contract NAS5-26555.
IA acknowledges support as a CIFAR Azrieli Global Scholar in the Gravity and the Extreme Universe Program, from the European Research Council (ERC) under the European Union’s Horizon 2020 research and innovation program (grant 852097), from the Israel Science Foundation (grant 2752/19), from the United States-Israel Binational Science Foundation (BSF), and from the Israeli Council for Higher Education Alon Fellowship.
BT acknowledges support from the Israel Science Foundation (grant 1849/19).
KAA is supported by the Danish National Research Foundation (DNRF132). 
JLD is supported by the GRF grant from the Hong Kong government under HKU 27305119. 
KDF is supported by Hubble Fellowship grant HST-HF2-51391.001-A from STScI, operated by AURA, Inc., under NASA contract NAS5-26555.
TW  is  funded  in  part  by  European Research  Council  grant  320360  and  by  European  Commission  grant  730980.
Parts of this research were supported by the Australian Research Council Centre of Excellence for All Sky Astrophysics in 3 Dimensions (ASTRO 3D), through project number CE170100013.
\end{acknowledgements}

%



\bibliographystyle{aps-nameyear}      

\bibliography{gen_update.bib,impostors.bib}
\end{document}